\documentclass[%
reprint,
 amsmath,amssymb,
 aps,
]{revtex4-1}

\usepackage{epsfig}
\usepackage{graphicx}
\usepackage{bm}
\usepackage{amssymb}
\usepackage{multirow}
\usepackage{dcolumn}

\emergencystretch=\maxdimen
\newcommand{\zh}{\bm}

\newcommand{\dee}{{\varepsilon}}

\newcommand{\zhr}{{\zh r}}

\newcommand{\zhe}{{\zh e}}

\newcommand{\zhk}{{\zh k}}
\newcommand{\zhalpha}{{\zh\alpha}}

\newcommand{\zhepsilon}{{\zh\epsilon}}
\newcommand{\zhep}{{\zh\epsilon}}

\newcommand{\zhA}{{\zh A}}
\newcommand{\zhY}{{\zh Y}}

\newcommand{\Br}[1]{(\ref{#1})}
\newcommand{\Eq}[1]{Eq.\ (\ref{#1})}

\newcommand{\Fig}[1]{Fig.\ \ref{#1}}
\newcommand{\txt}[1]{{\rm #1}}

\newcommand{\fun}{{\xi}}
\newcommand{\Fun}{{\Xi}}
\newcommand{\asym}{{a}}
\newcommand{\Asym}{{A}}

\newcommand{\vectorc}[3]{
\left(
\begin{array}{c}
#1\\
#2\\
#3\\
\end{array}
\right)
        }

\begin{document}

\title{Investigation of two-photon $2s\to 1s$ decay in one-electron and one-muon ions}

\author{V.\ A.\ Knyazeva}
\affiliation{St.\ Petersburg State University, 7/9 Universitetskaya nab.,
             St. Petersburg, 199034, Russia}
\author{K.\ N.\ Lyashchenko}
\affiliation{Institute of Modern Physics, Chinese Academy of Sciences, Lanzhou 730000, China}
\affiliation{Petersburg Nuclear Physics Institute named by B.P. Konstantinov of National
Research Centre “Kurchatov Institute”, Gatchina, Leningrad District 188300, Russia}

\author{M. Zhang}
\affiliation{Institute of Modern Physics, Chinese Academy of Sciences, Lanzhou 730000, China}
\affiliation{University of Chinese Academy of Sciences, Beijing 100049, China}

\author{D.\  Yu}
\affiliation{Institute of Modern Physics, Chinese Academy of Sciences, Lanzhou 730000, China}
\affiliation{University of Chinese Academy of Sciences, Beijing 100049, China}
\author{O.\ Yu.\ Andreev}
\affiliation{St.\ Petersburg State University, 7/9 Universitetskaya nab.,
             St. Petersburg, 199034, Russia}
\affiliation{Petersburg Nuclear Physics Institute named by B.P. Konstantinov of National
Research Centre “Kurchatov Institute”, Gatchina, Leningrad District 188300, Russia}

\date{\today}

\begin{abstract}
We have studied the radiative decay of the $2s$ state of one-electron and one-muon ions,
where the two-photon mechanism plays an important role.
Due to the nuclear size corrections the radiative decay of the $2s$ state in the electron and muon ions is qualitatively different.
Based on the accurate relativistic calculation, we introduced a two-parameter approximation, which makes it possible to describe the two-photon angular-differential transition probability for the polarized emitted photons with high accuracy.
The emission of photons with linear and circular polarizations was studied separately.
We also investigated the transition probabilities for the polarized initial and final states.
The investigation was performed for ions with atomic numbers $1 \le Z \le 120$.
\end{abstract}

\maketitle

\section{Introduction}
The two-photon transitions represent one of the fundamental processes in the atomic physics.
The two-photon decay is best studied for one-electron ions, which  is the dominant decay channel of the $2s$-electron state  for light and middle $Z$ H-like ions, where $Z$ is the atomic number.
The probabilities of one- and two-photon transitions become comparable for $Z\approx 40$.
The decay of the $2s$-electron state has been extensively studied in theoretical \cite{goppertmayer31,breit1940,Spitzer1951,klarsfeld1969382,johnson72,au1976,goldman1981,santos1998,labzowsky2005jpb38-265,surzhykov2005PhysRevA.71.022509,surzhykov2009,PhysRevA.93.012517}
and experimental works
\cite{Lipeles1965,Prior1972,Kocher1972,Marrus1972,Cocke1974,Kruger1975,Hinds1978,Gould1983,Dunford1989,Cheng1993,Dunford1993,Mokler_2004}.
In the reverse process, the two-photon excitation $1s \rightarrow 2s$, a record accuracy of measurement of the transition frequency in hydrogen was obtained \cite{PhysRevLett.110.230801}.
For the one-muon ions the two-photon decay is the main radiative channel for all ions.

The experimental investigation of the $2s\to1s$ transition in muon ions was performed in
\cite{MISSIMER1985179,pohl2010}.
Since significant progress has recently been made in the quality of muon beams
\cite{bogomilov2020},
the study of muon ions becomes relevant.

Unlike the one-photon decay, the emission spectrum of the two-photon decay has a continuous distribution determined by the energy conservation law. 
The study of the differential transition probabilities is of particular interest.
The energy-differential transition probabilities were investigated theoretically in \cite{Spitzer1951,klarsfeld1969382,johnson72,goldman81,Tung1984,santos1998,surzhykov2009,Sommerfeldt2020}.

For the one-electron ions the radiative corrections to the transition probabilities are investigated in works \cite{Karshenboim1997,Jentschura2004,Sommerfeldt2020}.
The dominant part of the electron self-energy radiative correction to the two-photon transition probabilities were calculated in
\cite{Karshenboim1997,Jentschura2004}.
The vacuum polarization corrections (in the Uehling approximation) are presented in
\cite{Sommerfeldt2020}.
Contribution of the negative continuum of the Dirac spectrum to the total and differential transition probabilities is investigated in
\cite{labzowsky2005jpb38-265}
and
\cite{surzhykov2009},
respectively.

The angle-differential transition probabilities have a nontrivial dependence on the angle between the emitted photons.
The angular distribution of the emitted photons is determined by the dominant E1E1 transitions, which gives $1+\cos^2\theta$ distribution, where $\theta$ is the angle between the momenta of the emitted photons.
The deviation from this distribution was investigated by Au in the nonrelativistic limit \cite{au1976}. 
The deviation leads to an asymmetry of the angle-differential transition probability, which is explained by the interference between the E1E1 and higher multipoles (mainly E2E2 and M1M1).
For the one-electron ions the asymmetry of the angular distribution was investigated 
for unpolarized photons emitted by ions of xenon and uranium in
\cite{surzhykov2005PhysRevA.71.022509}.

In this work, the asymmetry of the angular distribution is investigated for both unpolarized and polarized emitted photons for all one-electron and one-muon ions including the super heavy elements. 
The difference between the relativistic calculation of the asymmetry and the calculation of Au
\cite{au1976}
reaches three times for the super heavy elements. 
In the case of light ions, the asymmetry is small, but important for evaluating the nonresonant corrections
\cite{andreev08pr}.
The nonresonant corrections set the limit of the concept of energy levels and are already taken into account in the most accurate experiments
\cite{PhysRevLett.119.263002}.
The polarization properties of the two-photon transitions were studied in the processes of elastic photon scattering on atoms
\cite{zon68,Roy1986,Manakov_2000}.

We consider the two-photon decay of the $2s$ state of one-electron and one-muon ions with atomic numbers $1\le Z\le 120$ within the relativistic theory.
We found that the radiative decay of the $2s$ state in the electron and muon ions is qualitatively different.
In particular, for one-electron ions, the only cascade channel is $2s\to 2p_{1/2}\to 1s$, which is negligibly weak, mainly due to the small energy difference between $2s$ and $2p_{1/2 }$ states
\cite{Sommerfeldt2020}.
In the case of one-muon ions, the situation is different.
First, there is another cascade channel: $2s\to 2p_{3/2}\to 1s$.
Second, the energy difference between the $2s$ and $2p$ states is sufficiently large, so that the cascade channels become dominant already for the middle $Z$ ions.
All this radically changes the decay of the $2s$-muon state.

We also present the investigation of the angle differential transition probabilities with respect to the polarization of the emitted photons.
We  studied the differential transition probabilities for the emission of a photon with certain linear and circular polarizations, as well as the transition probabilities for polarized initial and final states.
Recently it was reported
that the detector technology for the measurement of the linear photon polarization (appearing in the K-shell radiative electron capture by heavy ions) was significantly improved
\cite{Vockert2019}.
We introduced a two-parameter approximation for the differential transition probabilities, which was used to analyze different polarizations of photons even in the relativistic domain.
The two-parameter approximation describes with a high accuracy the angle-differential transition probability (even for $Z=120$, the accuracy is better than $1\%$ for the photons with equal energies), in particular, it explicitly describes the asymmetry of the angular distribution.
It was found that the negative continuum of the Dirac equation is of great importance for the asymmetry parameter even for light ions (the transverse gauge was used).

\section{Theory}
In this paper, we consider the radiative decay of one-electron and one-muon ions.
Since the muon can be considered as a 'heavy' electron (the muon mass is about $207$ of the electron mass), the application of the theory developed for electron ions to muon ions consists in replacing the mass of the electron with the mass of the muon
\cite{RevModPhys.54.67}.
In this section, we present the theory of two-photon decay of one-electron ions.

We note, that we do not consider the magnetic hyperfine structure.
The hyperfine structure of the muon ions was investigated in
\cite{PhysRevC.2.102}. 

\begin{center}
\begin{figure}[h]
\includegraphics[width=0.35\textwidth]{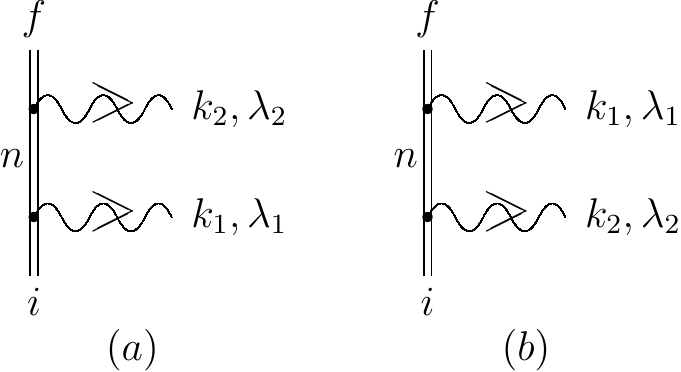}
\caption{Feynman graphs describing a two-photon transition in a one-electron ion.
The double solid lines denote an electron in the potential of the atomic nucleus (the Furry picture).
The wavy lines with the arrows describe the emission of a photon with momentum $k^{\mu}=(\omega,\zhk)$ and polarization $\lambda$.}
\label{feyn1}
\end{figure}
\end{center}

The two-photon decay of the $2s$ state of one-electron ions can be schematically depicted as
\begin{equation}
2s\rightarrow 1s+\gamma_1+\gamma_2\,.
\end{equation}
The Feynman graphs corresponding to the two-photon decay are presented in Fig.~\ref{feyn1}, where the double lines represent electrons in the electric field of the atomic nucleus (the Furry picture).
The graphs $(a)$ and $(b)$ in Fig.~\ref{feyn1} differ in the order of the emitted photons.
Index $n$ denotes the summation over the complete Dirac spectrum, including the positive and negative energy continuum. 

Using the Feynman rules, the S-matrix element for transition from the initial state ($i$) to the final state ($f$) corresponding to graph $(a)$ in Fig.~\ref{feyn1} can be written as
\begin{eqnarray}
S^{(a)}_{i\to f}
&=&\nonumber
(-ie)^2\int d^4 x_1 d^4 x_2 \overline{\psi}_f(x_2)\gamma^{\mu_2}A^{*(k_2,\lambda_2)}_{\mu_2}(x_2)
\\
&&
\times
S(x_2,x_1)\gamma^{\mu_1}A^{*(k_1,\lambda_1)}_{\mu_1}(x_1)\psi_i(x_1)\,,
\label{sif}
\end{eqnarray}
where $e$ is the electron charge,
\begin{equation}
S(x_1,x_2)=\frac{i}{2\pi}\int_{-\infty}^{\infty} d\omega_n e^{-i\omega_n(t_1-t_2)}\sum_{n}\frac{\psi_n(\zhr_1)\overline{\psi}_n(\zhr_2)}{\omega_n-\varepsilon_n(1-i0)}
\label{sx1x2}
\end{equation}
is the electron propagator, $\psi_i$ and $\psi_f$ are the wave functions of the initial and final states, respectively, $A_\mu$ is electromagnetic four-potential. 
 The sum includes the summation of the discrete Dirac spectrum and the integration over the positive and negative continuum.
 In \Eq{sx1x2} index $n$ denotes a set of quantum numbers ($n=(n,j,l,m)$) defining an intermediate state with principal quantum number $n$, angular momentum $j$, parity $(-1)^l$ and projection of the angular momentum $m$.
The photon wave functions $A^{\mu(k,\lambda)}$ are considered in the transverse gauge where the scalar photons are absent
\begin{equation}
A^{\mu(k,\lambda)}(x)=(0,\zhA^{(k,\lambda)}(\zhr,t))\,.
\label{Atr}
\end{equation}
The vector part of the photon 4-vector expresses as
\begin{equation}
\zhA^{(k,\lambda)}(\zhr,t)=\zhA^{(k,\lambda)}(\zhr)\mathrm{e} ^{i\omega t}\,.
\end{equation}
The relativistic units are used throughout the paper $(\hbar=1,c=1)$.

The amplitude is connected with the S-matrix as
\begin{eqnarray}
S_{i\to f}=-2\pi i \delta (\varepsilon_f+\omega_1+\omega_2-\varepsilon_i) U_{fi}\,,
\end{eqnarray}
where $\varepsilon_i$ and $\varepsilon_f $ are the energies of the initial and final states.

Integrating over the time variables in \Eq{sif} and introducing the following matrix elements
\begin{eqnarray}
A^{*(k_1,\lambda_1)}_{ni}&=&\int d\zhr \psi^+_n(\zhr)(-1)\zhalpha\zhA^{*(k_1,\lambda_1)}(\zhr)\psi_i (\zhr) \,, 
\label{a1} \\
A^{*(k_2,\lambda_2)}_{fn}&=&\int d\zhr \psi^+_f(\zhr)(-1)\zhalpha\zhA^{*(k_2,\lambda_2)}(\zhr)\psi_n (\zhr)\,,
\label{a2}
\end{eqnarray}
where $\zhalpha$ are the Dirac alpha-matrices,
we obtain the following expression for the amplitude corresponding to the graphs (a)
\begin{eqnarray}
U^{(a)}_{fi}
&=&\label{u1}
e^2 \sum_n \frac{A^{*(k_2,\lambda_2)}_{fn}A^{*(k_1,\lambda_1)}_{ni}}{\varepsilon_i-\omega_1-\varepsilon_n}\,, 
\end{eqnarray}
Similarly the expression for the graph $(b)$ in Fig.~\ref{feyn1} reads as
\begin{eqnarray}
U^{(b)}_{fi}
&=&\label{ub} 
e^2 \sum_n \frac{A^{*(k_1,\lambda_1)}_{fn}A^{*(k_2,\lambda_2)}_{ni}}{\varepsilon_i-\omega_2-\varepsilon_n}
\,.
\end{eqnarray}
In the case of the one-electron ions,
the energy of the intermediate $2p_{1/2}$ state is placed between the energies of $1s$ and $2s$ states ($\varepsilon_{1s}<\varepsilon_{2p_{1/2}}< \varepsilon_{2s}$),
the considered two-photon decay can proceed through the formation of the $2p_{1/2}$ state (cascade decay). Direct accounting for the $2p_{1/2}$ state leads to a zero denominator, if frequency of one of the photons is $\omega_{1,2}=\varepsilon_i-\varepsilon_{2p_{1/2}}$.  Considering this state, it is necessary to make numerous insertions of the electron self-energy Feynman graph into the internal electron line
\cite{low52}.
This procedure adds the self-energy correction $\Delta \varepsilon_{2p {1/2}} = L- \frac{i}{2} \Gamma $ to the Dirac energy $ \varepsilon_n $ corresponding to the $ 2p_{1/2} $ state, where $ L $ is the Lamb shift and $ \Gamma $ is the one-photon radiative width of the $ 2p_{1/2} $ state. We note that evaluation of the Lamb shift $ L $ is a question of renormalization and it is neglected in the present calculation.
However, the imaginary part of the self-energy correction (radiative width $\Gamma$) was taken into account. In this approach, zero denominators do not arise.

We note that in the case of one-muon ions, between the energies of the $1s$ and $2s$ states there is also the $2p_{3/2}$ state ($\varepsilon_{1s }<\varepsilon_{2p_{3/2}}< \varepsilon_ {2s}$).
Accordingly, such a procedure must be performed for the $2p_{3/2}$ state as well.

The total transition amplitude is the sum of the
contributions of the graphs (a) and (b) 
\begin{eqnarray}
U_{fi}=U^{(a)}_{fi}+U^{(b)}_{fi}\,.
\end{eqnarray}
The two-photon differential transition probability reads as
\begin{eqnarray}
dW^{(\lambda_1,\lambda_2)}_{fi}
&=& \nonumber
2\pi\Big|U_{fi}\Big|^2 \delta (\varepsilon_i-\omega_1-\omega_2-\varepsilon_f) \\
&&
\times
\frac{d\zhk_1}{(2\pi)^3}
\frac{d\zhk_2}{(2\pi)^3}\,.
\end{eqnarray}
After integration over one of the photon energies we obtain the following differential transition probability
\begin{eqnarray}
\frac{dW^{(\lambda_1,\lambda_2)}_{fi}}{d\Omega_1 d\Omega_2 d\omega_1}
&=& 
\frac{1}{(2\pi)^5}
\Big|U_{fi}\Big|^2 \omega_1^2\omega_2^2\,,
\end{eqnarray}
where $\Omega_{1,2}$ is the solid angle of the corresponding photon momentum.
The energy of the second emitted photon is determined by the energy conservation law
\begin{eqnarray} 
\omega_1+\omega_2=\varepsilon_i-\varepsilon_f. 
\label{enlaw}
\end{eqnarray}
It is convenient to introduce the energy sharing fraction
\begin{eqnarray} 
x(\omega_1)
&=&\label{eqn220211n01}
\frac{\omega_1}{\dee_i-\dee_f}
\,.
\end{eqnarray}

To describe the polarization of a photon, we introduce a unit vector  directed along the photon momentum
\begin{eqnarray}
\hat{\zhk}=
\begin{pmatrix}
\sin\theta_k \cos\varphi_k \\
\sin\theta_k \sin\varphi_k \\
\cos \theta_k
\end{pmatrix}\,,
\end{eqnarray}
the vector $\zhe_z=(0,0,1)$ and 
two vectors orthogonal to $\hat{\zhk}$
\begin{eqnarray}
\zhe^{(1)}=\frac{[\zhe_z\times\hat{\zhk}]}{|[\zhe_z\times\hat{\zhk}]|}\,, \hspace{0.5cm} 
\zhe^{(2)}=\frac{[\zhe^{(1)}\times\hat{\zhk}]}{|[\zhe^{(1)}\times\hat{\zhk}]|}\,.
\end{eqnarray}
In spherical coordinates, these vectors read as
\begin{equation}
\zhe^{(1)}=
\begin{pmatrix}
-\sin\varphi_k \\
\cos\varphi_k \\
0
\end{pmatrix}\,,\hspace{0.3cm}
\zhe^{(2)}=
\begin{pmatrix}
\cos\theta_k\cos\varphi_k \\
\cos\theta_k\sin\varphi_k \\
-\sin\theta_k
\end{pmatrix}\,.
\label{18}
\end{equation}
The photon polarization vector $\zhepsilon^{(\lambda)}$ can be presented as a linear combination of the vectors $\zhe^{(1)}$ and $\zhe^{(2)}$
\begin{eqnarray}
\zhepsilon^{(\lambda)} &=& \alpha_1\zhe^{(1)}+\alpha_2\zhe^{(2)}\,,
\end{eqnarray}
where $|\alpha_1|^2+|\alpha_2|^2=1$\,.

To calculate the matrix elements in Eqs.~(\ref{a1}) and (\ref{a2}), we use the partial wave expansion of the photon function

\begin{eqnarray}
\zhA^{(k,\lambda)*}(\zhr)
&=&\nonumber
\sqrt{\frac{2\pi}{\omega}}\zhepsilon^{(\lambda)*} e^{-i\zhk\zhr}
\\
&=&\nonumber
\sqrt{\frac{2\pi}{\omega}}\sum_{jlm}(-i)^l (\zhepsilon^{(\lambda)},\zhY_{jlm}(\hat{\zhk}))\\
&&
\times
4\pi j_l(kr)\zhY^*_{jlm}(\hat{\zhr})\,,
\end{eqnarray}
where $j_{}(kr)$ is the spherical Bessel function.
The scalar product involves the complex conjugation for the first element.
Integration over the spatial angular variables ($\hat{\zhr}$) is performed analytically, and integration over the radial variables is performed numerically.

In
Eqs.~(\ref{u1}) and (\ref{ub}),
summation over the complete Dirac spectrum is performed using the finite basis sets for the Dirac equation \cite{johnson88p307,shabaev04}.

We used the Fermi distribution of the nuclear charge density.
The nuclear root mean square charge radii were taken from \cite{Angeli2004,Kozhedub2008,Zhou2017}, they are listed in
Tables~\ref{tab-el-rad} and \ref{tab-muon-rad}.
We note, that for $Z = 1$ the
Fermi distribution is inapplicable, so we used the model
of a homogeneously charged sphere.
In the case of muon ions, the transition probabilities are sensitive to the nuclear model used.
To study it, we also performed a calculation with the model of a homogeneously charged sphere.

For the one-muon ions, 
the nuclear recoil correction was taken into account using the reduced muon mass.
In the case of the  muon ions, the main radiative correction is the electron vacuum polarization correction, which can be taken into account within the Uehling approximation.
For one-electron ions, 
the nuclear recoil correction and the radiative corrections were neglected
because of their smallness
\cite{Sommerfeldt2020}.

\section{Results and discussion}
\subsection{Transition probabilities}
We investigate the radiative decay of the $2s$ state in one-electron and one-muon ions.
We consider ions with atomic nuclear charges in the range from $1$ to $120$.
The main attention is paid to the role of two-photon decay.

The $2s$ state is the longest-lived among the states of the L-shell, i.e. the $2s$, $2p_{1/2}$, $2p_{3/2}$ states.
Tables~\ref{tab-el-2p2p2s} and \ref{tab-mu-2p2p2s}
give the transition probabilities for these states for electron and muon ions, respectively.
The radiative decay of $2p_{1/2}$, $2p_{3/2}$ states is determined by the one-photon ($E1$) transitions to the $1s$ state.
For the electron ions, the dominant channel of the $2s$ state radiative decay depends on the nuclear charge $Z$: for light ions, the two-photon (mainly $E1E1$) transitions predominate, while for heavier ions ($Z>40$), the decay is determined by one-photon ($M1$) transitions.
For muon ions the radiative decay is determined by the two-photon ($E1E1$) transition for all $Z$.
Therefore, in
Tables~\ref{tab-el-2p2p2s} and \ref{tab-mu-2p2p2s},
the total $2s\to1s$ transition probabilities  are given as the sum of the one- and two-photon transition probabilities.
Below we consider the decay of $2s$ states in more detail. 
The last column of
Table~\ref{tab-el-2p2p2s}
contains data for the $2s\to2p_{1/2}$ transition probability for one-electron ions.
The data show that this cascade channel is very small.
This is explained by the fact that the $2s$ and $2p_{1/2}$ energy levels are very close in one-electron ions. 
We will show that in the case of one-muon ions this channel is significant.

The results presented in
Tables~\ref{tab-el-2p2p2s} and \ref{tab-mu-2p2p2s}
are obtained separately for the point-like nucleus ($W^{(e,\mu)}_0$) and for the Fermi distribution of nuclear charge density ($W^{(e,\mu)}$).
The data show that for the electron ions, the nuclear size corrections are noticeable only for very heavy ions, while for the muon ions they are of great importance even for light ions.
For example, for the muon ions with $Z = 50$ the $2s \to 1s$ transition probabilities calculated with the point-like and the Fermi-distribution of nuclear charge density differ by one order of magnitude.
Another remarkable fact is that for muon ions the nuclear size corrections decrease the transition probabilities for the $2p_{1/2}$ and $2p_{3/2}$ states, but increase them for the $2s$ state for $Z>5$.

For the one-electron ions, the one-photon transition probabilities are listed in
\cite{jitrik2004},
the nuclear size corrections are considered in
\cite{popov2017}.
The two-photon transitions for the $2s$-electron state were investigated by many authors
\cite{au1976,labzowsky2005jpb38-265,surzhykov2009,PhysRevA.93.012517,Sommerfeldt2020}.

The results of calculating the transition probabilities for the $2s$ state of one-electron ions are presented in
Table~\ref{tab21}. 
In the double column marked `$\mbox{M1:}\,\,2s\to1s$' the one-photon (M1) transition probabilities ($W^{(e)}$) and their power dependence ($p_W$) on $Z$ are given.
For small $Z$ the transition probabilities are proportional to $Z^{10}$, for large $Z$ this dependence changes, reaching $Z^{13}$ for $Z=120$. 
The next multicolumn ($\mbox{E1E1:}\,\,2s\to1s$) presents the results for the two-photon transition probabilities ($W^{(e)}$) and their power dependence ($p_W$); the calculation was carried out in the approximation where only E1 photons were taken into account. 
In the columns marked  `$2s\to1s, \mbox{Total}\,\, 2\gamma$' the results for the total two-photon transition probability ($W^{(e)}$) and their power dependence ($p_W$) are listed. 
In the mentioned columns, the transition probabilities were obtained by averaging over the  projections of the total angular momentum of the initial state ($m_i$) and summing over the final projections ($m_f$).
Due to the different power dependence ($p_W$) on $Z$ of the one- and two-photon transition probabilities, the two-photon decay dominates for $Z\le 40$, while  for larger $Z$, the decay occurs mainly via single photon M1 emission.
The presented data show that taking into account only the E1 photons is a good approximation:
even for heavy elements its accuracy does not exceed $1\%$.
However, below we will show that higher multipoles are of importance for differential transition probabilities. 
The transition probabilities for non-spin-flip ($m_i = m_f$) and spin-flip ($m_i =- m_f$) transitions are listed separately in the next columns.
The spin-flip transition probability for one-photon M1 transition is $2/3\, W^{(e,\mu)}$, the non-spin-flip transition is $1/3\, W^{(e,\mu)}$, where $W^{(e,\mu)}$ is the total M1 transition probability either for a one-electron or one-muon ion, respectively.
In contrast to the one-photon transition, the two-photon transitions occur mainly without the spin-flip.
We note, that in the case of the two-photon transitions, the non-spin-flip and spin-flip have different $Z$ dependencies.
For a hydrogen atom, the two-photon transition probability with the non-spin-flip is $9$ orders of magnitude larger than the transition probability with the spin-flip.
For the heavy ions, this difference is only $1.5$ order of magnitude.
We also compare our results for transition probabilities with the results obtained in work \cite{labzowsky2005jpb38-265}, where only E1E1 transitions were considered.
Our results are in reasonable agreement.
The transition probabilities for a point-like nucleus are presented in work
\cite{PhysRevA.93.012517},
where an analytical expression for the Dirac Coulomb Green's function was used, and in work
\cite{Sommerfeldt2020},
where the finite basis sets for the Dirac equation constructed from \textit{B} splines was employed \cite{johnson88p307,shabaev04} (as in the present work).
In work \cite{Sommerfeldt2020} the vacuum polarization corrections (in the Uehling approximation) to the transition probabilities were calculated.
The results of our calculation (for $1\le Z\le 92$) are in complete agreement with work \cite{Sommerfeldt2020}, so we do not give their results.

\begin{figure}
\includegraphics[scale=0.35]{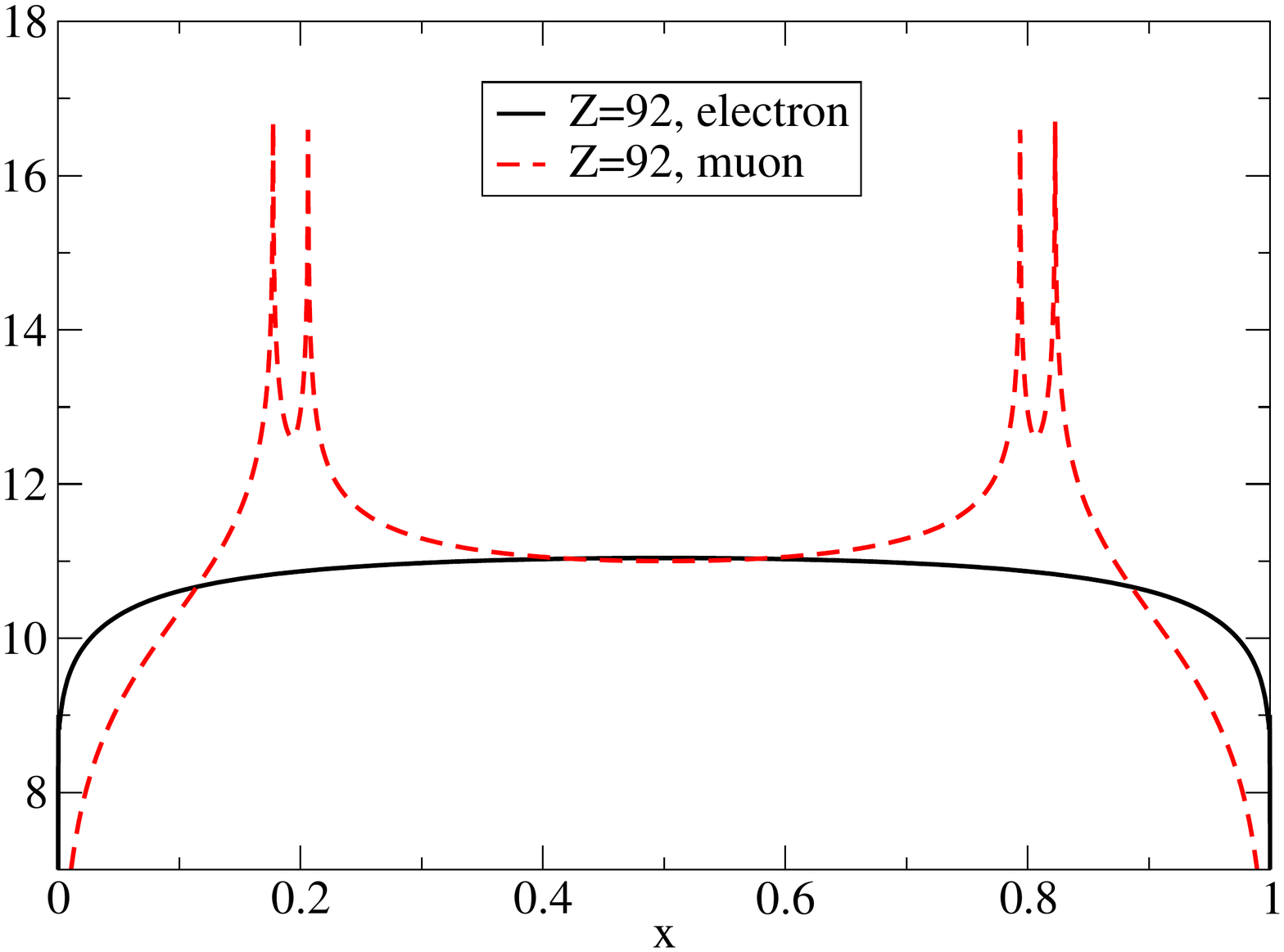}
\caption{
The differential transition probabilities (in $\mbox{s}^{-1} \mbox{keV}^{-1}$) for
electron and muon ions for $Z=92$ are presented as a function of the energy sharing fraction $x$
[see
\Eq{eqn220211n01}].
The differential transition probabilities are given on a logarithmic scale as
$\log_{10}\left(\frac{d W^{(e,\mu)}}{d\omega_1}\right)$.
}
\label{fig1-el-mu-1}
\end{figure}

\begin{figure}
\includegraphics[scale=0.35]{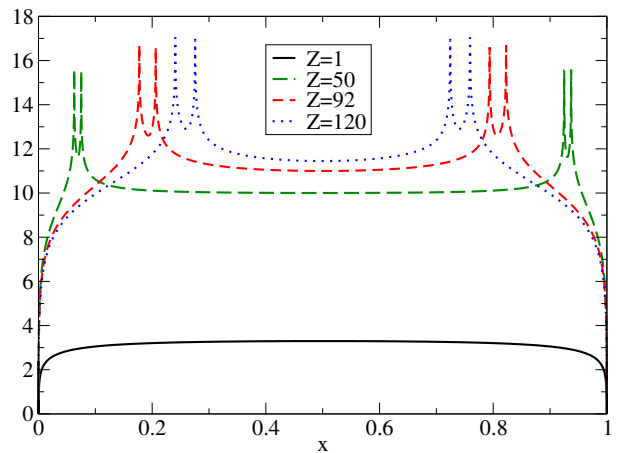}
\caption{
The differential transition probabilities for muon ions presented as a function of the energy sharing fraction $x$
[see
\Eq{eqn220211n01}].
The differential transition probabilities are given on a logarithmic scale as
$\log_{10}\left(\frac{d W^{(\mu)}}{d\omega_1}\right)$.
}
\label{fig1-mu-1}
\end{figure}

In
Tables~\ref{tab-tp-2s-muon} and \ref{tab-tp-2s-spinf-muon}
we give various transition probabilities for the radiative decay of the $2s$ state of the muon ions.
We see that the radiative decay channels of the $2s$ state for electron and muon ions are qualitatively different.
First of all, it should be noted that the order of the energy levels of the L-shell for electron ions and for muon ions is different. 
In
Tables~\ref{tab-el-rad} and \ref{tab-muon-rad}
we present the bound energies of the $1s$, $2s$, $2p_{1/2}$ and $2p_{3/2}$ states for one-electron ions and one-muon ions, respectively.
We can see that, in the case of one-electron ions, only the $2p_{1/2}$ state is placed between the $2s$ and $1s$ states
and the possible cascade channel of decay $2s\to2p_{1/2}$ is very week even for super heavy ions
(see the last column of
Table~\ref{tab-el-2p2p2s}).
In the case of one-muon ions the both $2p_{1/2}$ and $2p_{3/2}$ states are below the $2s$ state (see
Table~\ref{tab-muon-rad}).
The cascade decay channels $2s\to2p_{1/2}$ and   $2s\to2p_{3/2}$ are strong and become dominant for $Z>30$ (see
Table~\ref{tab-tp-2s-muon}).
In
Fig.~\ref{fig1-el-mu-1}
we present the differential transition probabilities for electron and muon uranium ions presented as a function of $x$
(the parameter $x$ is defined by \Eq{eqn220211n01}).
The figure demonstrates that for the electron ions the contribution of the cascade channel is not noticeable, while for the muon ions the cascade channels dominate.
The differential transition probabilities are symmetric with respect to the middle energy of the emitted photon ($x=\frac{1}{2}$).
The cascade transitions $2s\to2p_{1/2}\to1s$ (the first and the fourth peeks) and $2s\to2p_{3/2}\to1s$ (the second and the third peeks)  are represented by the resonances in the differential transition probabilities.
In
Fig.~\ref{fig1-mu-1}
we show the differential transition probabilities for the muon ions for several $Z$.
We can see the increase in the contribution of the cascade transitions with increasing nuclear charge $Z$.
Since the cascade channels are strong in muon ions, the energy of the $2s\to2p$ and $2p\to1s$ transitions can be measured, which will provide information about the structure of atomic nuclei.

\begin{figure}
\includegraphics[scale=0.35]{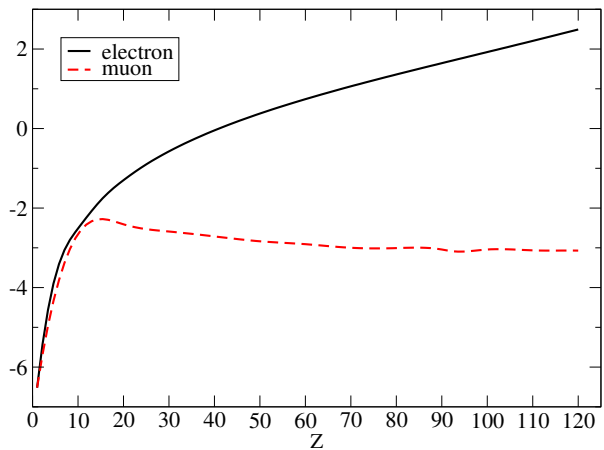}
\caption{
The ratio between the one-photon and two-photon transition probabilities for electron (black solid curve) and muon (red dashed line) ions presented as function of  the nuclear charge $Z$:
$\log_{10}\left(W^{(e,\mu)}_{\txt{1ph}}/W^{(e,\mu)}_{\txt{2ph}}\right)$.
}
\label{fig-ratio}
\end{figure}

The second important feature of the muon ions is that the nuclear size corrections are of great importance.
For $Z>5$
these corrections decrease the one-photon transition probabilities and increase the two-photon transition probabilities (see
Table~\ref{tab-tp-2s-muon}).
Since, in the case of the muon ions, the two-photon decay of the $2s$ state is dominant for all $Z$, the nuclear size correction increase the total transition probability of the $2s$ state.
The importance of the nuclear size correction for the muon ions is explained by the fact that the muon is placed much more close to the nucleus than the electron.
The values of the root mean square orbital radius of the corresponding states
($r^{(e,\mu)}=\langle\psi^{(e,\mu)}|r^2|\psi^{(e,\mu)}\rangle^{1/2}$) are given in
Table~\ref{tab-el-rad} and \ref{tab-muon-rad}.
We see that in the case of the muon ions the root mean square radii of the L-shell states are very close to the root mean square radii of the nuclei ($R$).
In
Fig.~\ref{fig-ratio}
we present the ratio between the one-photon and two-photon transition probabilities for the electron and muon ions.
For small $Z$ these ratios are close for electron and muon ions, but for heavy ions they become very different.
The difference between these ratios shows the role of the nuclear size corrections for the muon ions.

The nuclear size corrections, determined by the nuclear charge radii, are of great importance for the one-muon ions.
However, the radii of the nuclei depend on $Z$ non-linearly.
Accordingly, the $Z$-dependence of the transition probabilities (where the nuclear corrections are taken into account) is cumbersome.
So, we do not present the $Z$-dependence of the  transition probabilities for the one-muon ions where the nuclear charge corrections are taken into account.

In
Tables~\ref{tab-mu-2p2p2s}, \ref{tab-tp-2s-muon}
we compare our results with the data presented in \cite{MISSIMER1985179}.
In general, our data are in reasonable agreement.
The only serious discrepancy is found for the two-photon transition probability for $Z=30$ in
Table~\ref{tab-tp-2s-muon}.

We also investigated the contribution of the E1E1 transition for the muon ions and the separate contributions of the spin-flip and non-spin-flip transitions.
In
Table~\ref{tab-tp-2s-spinf-muon}
we can see that
for the muon ions as well as for the electron ions the E1E1 transition is dominant.
However, in contrast to the one-electron ions, for the one-muon ions the spin-flip transition becomes significant for $Z>10$.

For the one-electron ions, the nuclear size corrections and the vacuum polarization corrections (in the Uehling approximation) for the two-photon transition probabilities were investigated in
\cite{Sommerfeldt2020}.
In general, these corrections are noticeable only for the heavy ions.
In contrast to the one-electron ions, for the one-muon ions these corrections are of importance even for the light ions. 
In Table~\ref{tab-muon-tptp-details}
we present various corrections to the transition probabilities.
We see that the transition probabilities calculated with the point-like nucleus
have the same power dependence on $Z$ as the transition probabilities for the one-electron ions.
We can also see the importance of the nuclear size correction: the difference between the data for the point-like nucleus and the data obtained with the Fermi distribution for the nuclear charge density exceeds two orders of magnitude for the heavy ions.

The nuclear recoil correction is taken into account using the reduced muon mass.
This correction is important only for the light ions.
The vacuum polarization  correction was taken into account within the Uehling approximation.
For the one-muon ions the vacuum polarization correction is noticeable for ions with $Z>10$.

Since the nuclear size corrections are large for the one-muon ions,
we investigated the dependence of these corrections on the nuclear model.
In
Table~\ref{tab-muon-tptp-ball}
we present the results of calculation with two models of distribution of the nuclear charge density: the Fermi distribution and the nucleus considered as a homogeneously charged sphere.
We can see that the difference between these two models reaches $3\%$ for super heavy ions.
We estimate the accuracy of our calculation by the difference between these models.

\subsection{Two-parameter approximation}
\begin{figure}
\includegraphics[scale=0.35]{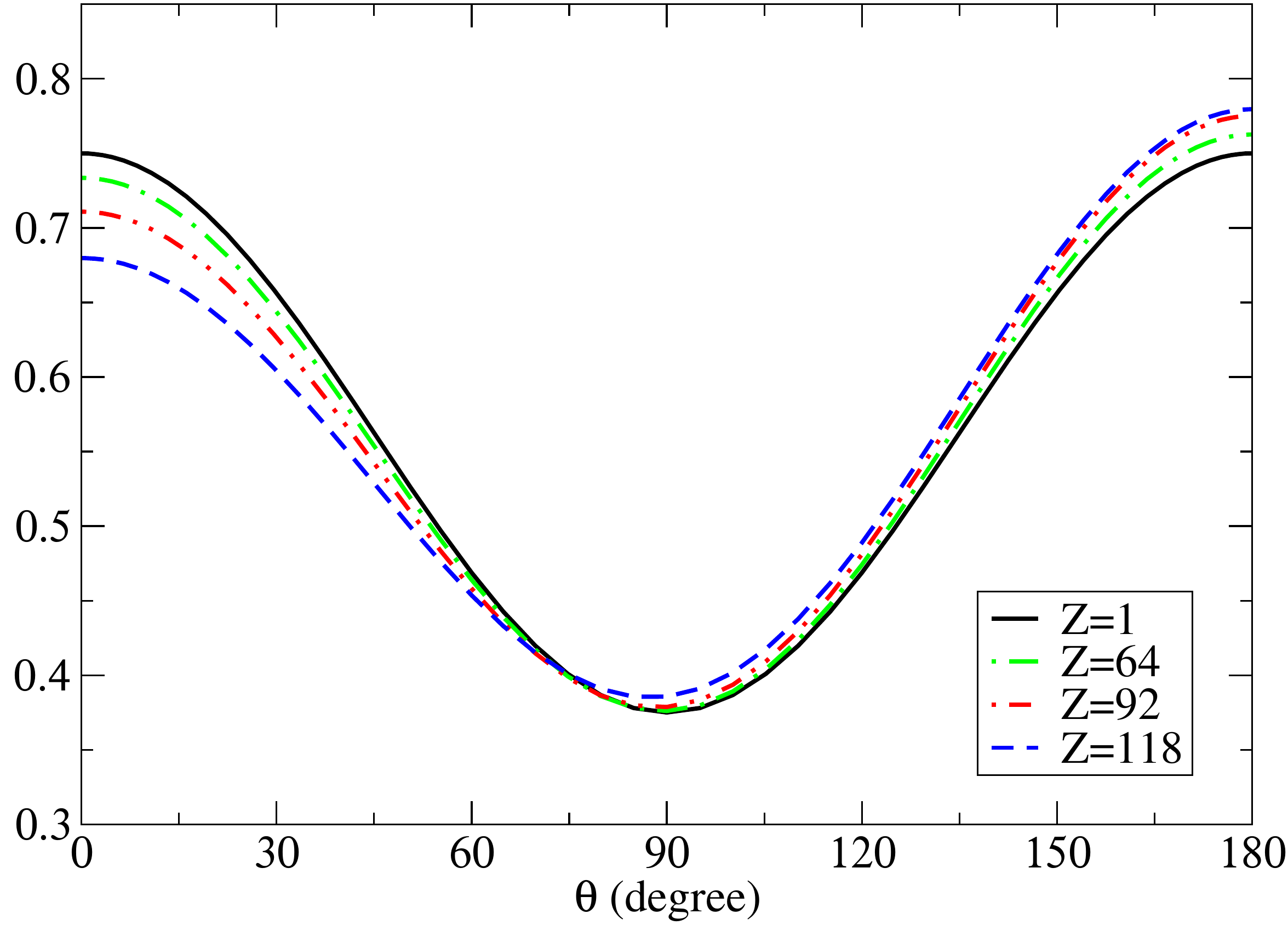}
\caption{
The normalized differential transition probabilities ($\frac{dW^{(e)}}{W^{(e)}\sin\theta d\theta }$) for one-electron ions as a function of the angle between the momenta of the emitted photons ($\theta$).
The data are presented for $Z=1, 64, 92, 118$.
}
\label{fig1}
\end{figure}

\begin{figure}
\includegraphics[scale=0.35]{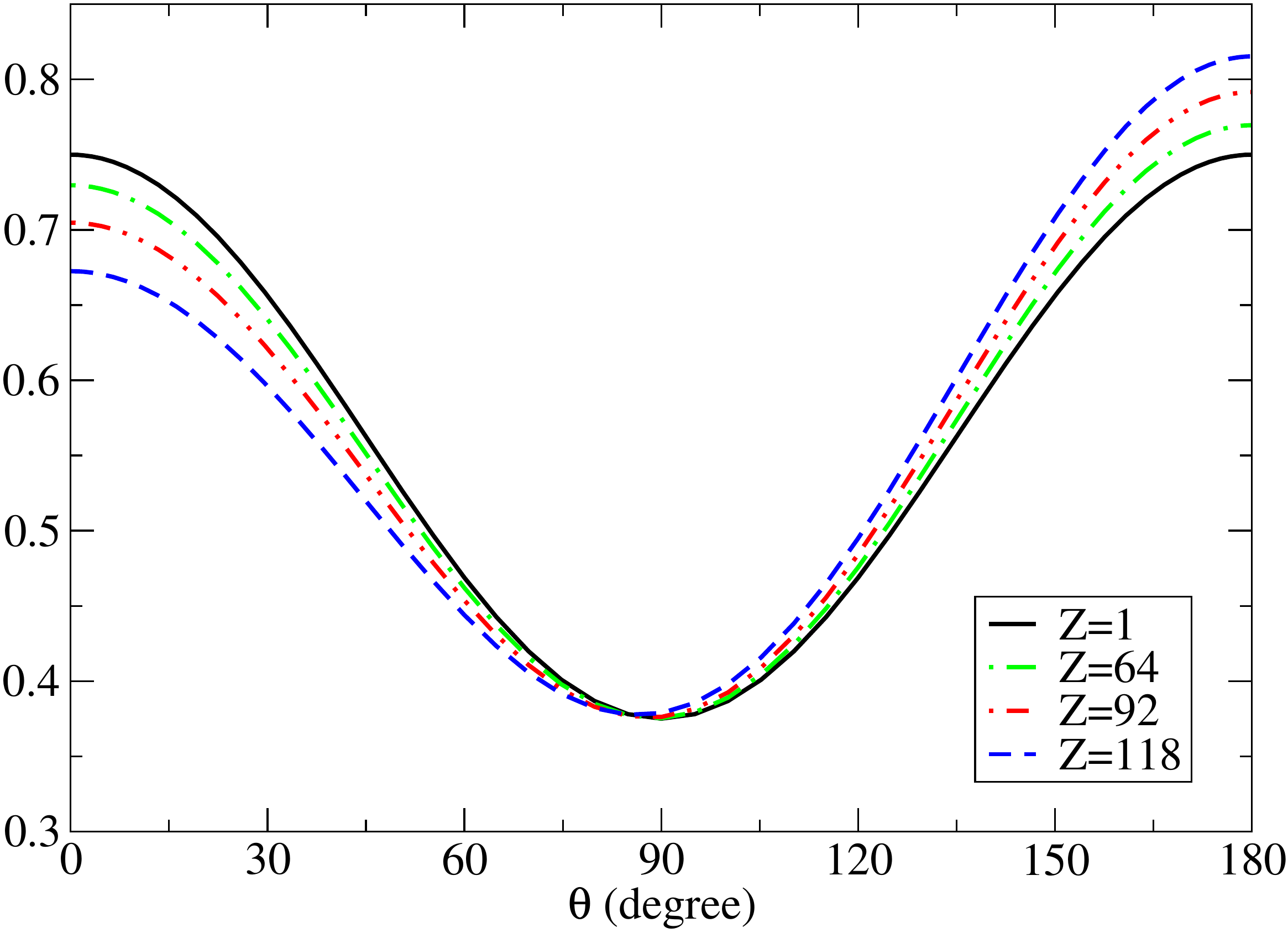}
\caption{The normalized differential transition probabilities
$\left[\left(\frac{dW^{(e)}}{\sin\theta d\theta d\omega_1}\right) \left(\frac{dW^{(e)}}{d\omega_1}\right)^{-1}\right]$
for one-electron ions
as a function of the angle between the momenta of the emitted photons ($\theta$) for $x=1/2$.
The data are presented for $Z=1,  64,  92,  118$.
}
\label{fig2}
\end{figure}

We performed calculation of the differential transition probability as a function of the angle between the momenta of the emitted photons ($\theta$).
The results of calculations of differential (over the angle $\theta$) transition probabilities for one-electron ions for several $Z$ are presented in
Fig.~\ref{fig1}.
The results for differential (over the angle $\theta$ and energy $\omega_1$) transition probabilities are given in Fig.~\ref{fig2}.
These results are in good agreement with work
\cite{surzhykov2005PhysRevA.71.022509}.
We found that the differential transition probability can be approximated with two parameters: the total two-photon transition probability $W$ and the asymmetry factor $\Asym$
\begin{eqnarray}
\frac{dW}{\sin\theta d\theta}
&=&\label{eqn210122n01}
\frac{3}{8}\left(1+\cos^2\theta\right)\Fun(\theta)W
\,,
\\
\Fun(\theta)
&=&\label{eqn210122n02t}
1-\Asym\cos\theta
\,.
\end{eqnarray}
The asymmetry factor is derived as
\begin{eqnarray}
A&=&\frac{\frac{dW}{\sin\theta d\theta}(180^\circ)-\frac{dW}{\sin\theta d\theta}(0^\circ)}{\frac{dW}{\sin\theta d\theta}(180^\circ)+\frac{dW}{\sin\theta d\theta}(0^\circ)} \,.
\label{A}
\end{eqnarray}

The angular distribution $1 + \cos^2 \theta$ is determined by the E1E1 transition \cite{au1976,Manakov_2000}.
The asymmetry of the angular distribution is explained by the interference between the E1E1 multipole and the M1M1 or E2E2 multipoles. The higher multipoles make a small contribution to the asymmetry even for super heavy ions.
The differential (over the angle $\theta$ and energy $\omega_1$) transition probability can also be approximated by two parameters: the differential (over energy of the emitted photon) transition probability $dW/d\omega_1$ and the asymmetry factor $a$
\begin{eqnarray}
\frac{dW}{\sin\theta d\theta d\omega_1}
&=&\label{eqn210121n01y}
\frac{3}{8}\left(1+\cos^2\theta\right)\fun(\theta,\omega_1)\frac{dW}{d\omega_1}
\,,
\\
\fun(\theta,\omega_1)
&=&\label{eqn210121n02}
1-\asym(x)\cos\theta
\,.
\end{eqnarray}
The asymmetry factor $a(x)$ is derived similarly to \Eq{A}.
The approximations (\ref{eqn210122n01})-(\ref{eqn210121n02}) for differential transition probabilities is called the two-parameter approximation.

The calculated values of the  transition probabilities $W$ and the asymmetry factors $\Asym$ are listed in
Table~\ref{tab1}
for one-electron ions and in
Table~\ref{tab-muon-tptp-details}
for one-muon ions.
We can see that for the muon ions, taking into account the  nuclear size corrections leads to a significant decrease in the asymmetry factors.
For the small $Z$ ions, the asymmetry factors for muon and electron ions are comparable, while for middle and heavy ions, the asymmetry factors for muon ions are $3$-$4$ orders of magnitude smaller than for an electron ion.
Accordingly, we will focus on the study of the asymmetry of the angular distribution of the emitted photon only for one-electron ions, where it is significant.
However, for the light ions, despite their small values, nonzero asymmetry factors lead to the appearance of nonresonant corrections, which are discussed in
Subsection~\ref{Nonresonant_corrections}.

In
Table~\ref{tab1}, we also give the asymmetry factor obtained from the nonrelativistic calculations \cite{au1976}.
Our results show good agreement with \cite{au1976} for light ions.
However, the discrepancy between our results for $Z=50$ is about $5\%$, for $Z=92$ -- $41\%$ and for $Z=118$ -- more than 3 times.
For middle $Z$ and heavy ions the results of nonrelativistic calculation \cite{au1976} are larger than our results.
The calculated values for the differential transition probabilities $dW/d\omega_1$ with the corresponding asymmetry factors $\asym(x)$ for $x=1/2, 1/3$ and $ 1/6$ are presented in
Tables~\ref{tab2}-\ref{tab4}.
The accuracy of the parameters $\Asym$ and $\asym(x)$ is determined by the accuracy of the approximations
\Br{eqn210122n01}
and
\Br{eqn210121n01y},
respectively.
The best accuracy of the two-parameter approximation reaches for $x=1/2$.
This accuracy is better than $6\times 10^{-3}\%$ for $Z=1$ and becomes $1\%$ for $Z=120$.
In Table~\ref{tab2} ($x=1/2$) we also give the asymmetry factor derived from work \cite{surzhykov2005PhysRevA.71.022509}.
In work \cite{surzhykov2005PhysRevA.71.022509} the calculations were carried out for point nuclei, which explains the difference between our results.
As follows from Tables~\ref{tab2}-\ref{tab4}, the asymmetry factor $\asym(x)$ depends on the energy of the emitted photon ($\omega_1$, $\omega_2$). 

However, the asymmetry factors $\Asym$ and $\asym(x)$ are almost independent of the angle between the emitted photons.

\subsection{Photon polarizations}

To study different polarizations of the emitted photons, it is convenient to employ the two-parameter approximation.

We consider the two-photon emission in coplanar geometry.
We assume that the momenta of the emitted photons are placed in the $(x,y)$-plane, i.e., the polar angles of the photon momenta are $\theta_1=\theta_2=\pi/2$.
The $x$-axis is directed along the momentum of the first photon $\zhk_1=\omega_1\zhe_{x}$\,.
In this case, the angle between the momenta of the emitted photons is determined by the azimuth angle of the second photon: 
$\theta=\min(\varphi_2,2\pi-\varphi_2)$\,.
Then the vectors $\zhe_1, \zhe_2$ (introduced in \Eq{18}) for the first photon read as
\begin{eqnarray}
\zhe^{(1)}_1
&=&\label{eqn210122n05}
\vectorc{0}{1}{0}
\,,\qquad
\zhe^{(2)}_1
\,=\,
\vectorc{0}{0}{-1}
\label{27}
\end{eqnarray}
and ones of the second photon are
\begin{eqnarray}
\zhe^{(1)}_2
&=&\label{eqn210122n06}
\vectorc{-\sin\varphi_2}{\cos\varphi_2}{0}
\,,\qquad
\zhe^{(2)}_2
\,=\,
\vectorc{0}{0}{-1}
\,.
\label{28}
\end{eqnarray}

Since all the plates composed by the emitted photons momenta are equivalent, the differential transition probability can be written as 
\begin{eqnarray}
\frac{dW^{(\lambda_1 \lambda_2)}}{\sin\theta d\theta d\omega_1}(\theta)
&=&
8\pi^2
\frac{dW^{(\lambda_1 \lambda_2)}}{d\Omega_1 d\Omega_2 d\omega_1}\Big(\frac{\pi}{2},0,\frac{\pi}{2},\theta\Big)
\,.
\end{eqnarray}
Then the energy-differential transition probability and the total transition probability read as
\begin{eqnarray}
\frac{dW}{d\omega_1} &=& \sum\limits_{\lambda_1 \lambda_2}\int\limits^{\pi}_{0} d\theta \sin\theta  \frac{dW^{(\lambda_1 \lambda_2)}}{\sin\theta d\theta d\omega_1}\,,
\\
W &=& \sum\limits_{\lambda_1 \lambda_2}\int\limits^{\pi}_{0} d\theta \sin\theta \int\limits^{(\varepsilon_i-\varepsilon_f)/2}_{0} d\omega_1 \frac{dW^{(\lambda_1 \lambda_2)}}{\sin\theta d\theta d\omega_1}\,.
\end{eqnarray}
Within the dipole approximation, the differential transition probability is proportional to \cite{zon68, au1976, Manakov_2000}
\begin{eqnarray}
\frac{dW^{(\lambda_1 \lambda_2)}}{d\Omega_1 d\Omega_2 d\omega_1}(\theta_1,\varphi_1,\theta_2,\varphi_2) \nonumber
\\
&&\hspace{-50pt}\sim\,
|\zhep^{(\lambda_1)}_{1}(\theta_1,\varphi_1)
\label{eqn210122n07x}
\cdot\zhep^{(\lambda_2)}_{2}(\theta_2,\varphi_2)|^2
\,.
\end{eqnarray}
Within the two-parameter approximation the differential transition probability for the polarized emitted photons reads as

\begin{eqnarray}
\frac{dW^{(\lambda_1 \lambda_2)}}{\sin\theta d\theta d\omega_1}
&=&\label{eqn210121n01x}
\frac{3}{8}|\zhep^{(\lambda_1)}_{1}(\theta_1,\varphi_1)
\label{eqn210122n07}\nonumber
\cdot\zhep^{(\lambda_2)}_{2}(\theta_2,\varphi_2)|^2\\
&&
\times
\fun(\theta,\omega_1)\frac{dW}{d\omega_1}
\,.
\end{eqnarray}

Below we consider three different polarizations of the emitted photons.

\subsubsection{Linear polarization $\zhep^{(0^{\circ})}$, $\zhep^{(90^{\circ})}$}
\label{lp1}

The polarization vectors of the first ($i=1$) and second ($i=2$) photons are chosen as
$\zhep^{(0^{\circ})}_i=\zhe^{(1)}_i$ and $\zhep^{(90^{\circ})}_i=\zhe^{(2)}_i$, respectively.
In this case, the polarization vectors $\zhep^{(0^{\circ})}_i$ are placed in the $(x,y)$-plane.

\begin{figure*}
\includegraphics[scale=0.5]{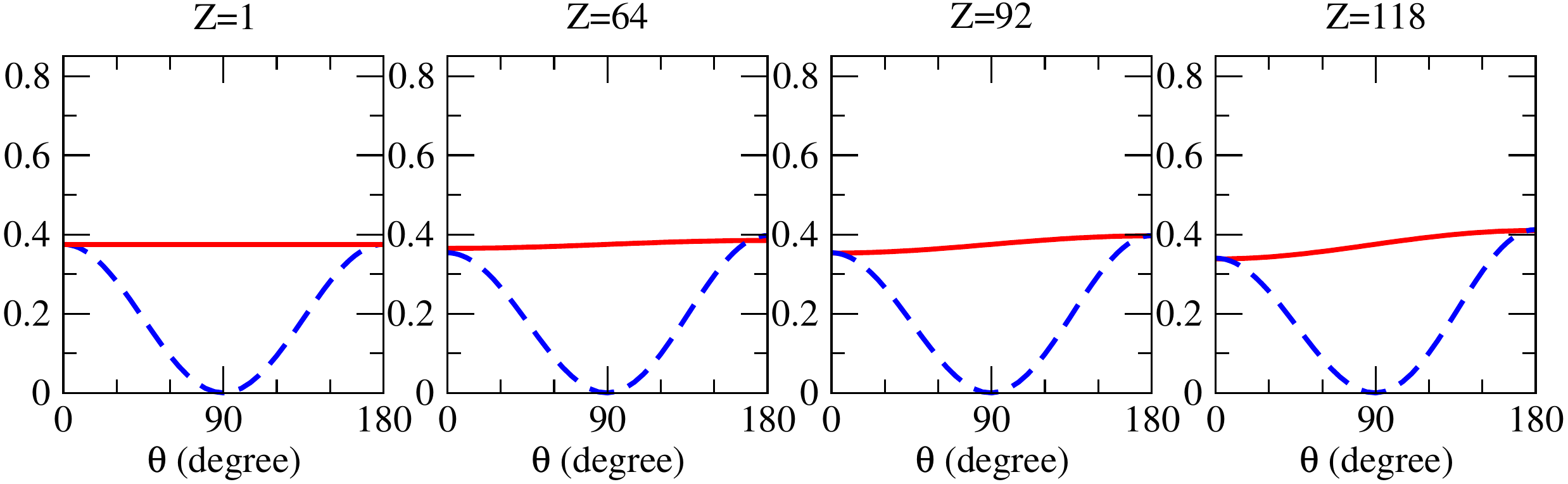}
\caption{The normalized differential transition probabilities
$\left[\left(\frac{dW^{(e)(\lambda_1 \lambda_2)}}{\sin\theta d\theta d\omega_1}\right) \left(\frac{dW^{(e)}}{d\omega_1}\right)^{-1}\right]$ as a function of the angle between the momenta of the emitted photons ($\theta$) for $x=1/2$.
The results for the photon linear polarizations $\zhep^{(0^{\circ})}$, $\zhep^{(90^{\circ})}$ considered in Subsection \ref{lp1} are presented.
The blue dashed line gives the angular dependence of the differential transition probability  for the  photons with polarizations $\zhep^{(90^{\circ})}_1$, $\zhep^{(90^{\circ})}_2$
[see
\Eq{34}].
The red solid line gives angular dependence of the differential transition probability for the  photons with polarizations $\zhep^{(0^{\circ})}_1$, $\zhep^{(0^{\circ})}_2$
[see
\Eq{33}].
The data are presented for $Z=1,  64,  92,  118$.
}
\label{fig3}
\end{figure*}

Accordingly, the differential transition probabilities for the considered photon linear polarization looks like
\begin{eqnarray}
\frac{dW^{(0^{\circ},0^{\circ})}}{\sin\theta d\theta d\omega_1}
&=& \label{33}
\frac{3}{8}\,\fun(\theta,\omega_1)\frac{dW}{d\omega_1}\,,
\\
\frac{dW^{(90^{\circ},90^{\circ})}}{\sin\theta d\theta d\omega_1} \label{34}
&=&
\frac{3}{8}\cos^2\theta \,\fun(\theta,\omega_1)\frac{dW}{d\omega_1}\,,
\\
\frac{dW^{(0^{\circ},90^{\circ})}}{\sin\theta d\theta d\omega_1}
&=&
\frac{dW^{(90^{\circ},0^{\circ})}}{\sin\theta d\theta d\omega_1}
\,=\,
0
\,.
\end{eqnarray}
The function $\fun(\theta,\omega_1)$ is given by
Eq.~\Br{eqn210121n02}.
The numerical results for the differential transition probabilities (for $x=1/2$) as a function of the angle $\theta$ are presented in
\Fig{fig3}.
The results of the exact numerical calculation and the results of the two-parameter approximation the are not distinguishable in this scale.
The blue dashed line represents the angular dependence of the differential transition probability \Eq{34}.
The red solid line gives the angular dependence of the differential transition probability \Eq{33}.
According to \Eq{33}, the red solid line shows the angular dependence of the function $\fun(\theta,\omega_1)$. 
In the case of this linear polarization, the contributions of photons with polarizations of $\zhep^{(0^{\circ})}_i$ and $\zhep^{(90^{\circ})}_i$ are very different.

\subsubsection{Linear polarization $\zhep^{(45^{\circ})}$, $\zhep^{(135^{\circ})}$}
\label{lp2}
In this Subsection we consider the polarization vectors chosen as
\begin{eqnarray}
\zhep^{(45^{\circ})}_i(\theta_i,\varphi_i)
&=&
\frac{1}{\sqrt{2}}(\zhe_i^{(1)}(\theta_i,\varphi_i)+\zhe_i^{(2)}(\theta_i,\varphi_i))\,,
\\
\zhep^{(135^{\circ})}_i(\theta_i,\varphi_i)
&=&
\frac{1}{\sqrt{2}}(\zhe_i^{(1)}(\theta_i,\varphi_i)-\zhe_i^{(2)}(\theta_i,\varphi_i))\,,
\end{eqnarray}
where $\zhe_i^{(1)}$, $\zhe_i^{(2)}$ are given by Eqs. (\ref{27}), (\ref{28}).
Index $i=1,2$ denotes the photon number.

Using \Eq{eqn210121n01x}, we can write the differential transition probabilities in the two-parameters approximation as 
\begin{eqnarray}
\frac{dW^{(45^{\circ},45^{\circ})}}{\sin\theta d\theta d\omega_1}
= \label{38}
\frac{dW^{(135^{\circ},135^{\circ})}}{\sin\theta d\theta d\omega_1}
&=&
\frac{3}{4}\cos^4\frac{\theta}{2} \nonumber \\
&&
\times
\fun(\theta,\omega_1)\frac{dW}{ d\omega_1}\,,
\\
\frac{dW^{(45^{\circ},135^{\circ})}}{\sin\theta d\theta d\omega_1}
= \label{39}
\frac{dW^{(135^{\circ},45^{\circ})}}{\sin\theta d\theta d\omega_1}
&=&
\frac{3}{4}\sin^4\frac{\theta}{2} \nonumber \\
&&
\times
\fun(\theta,\omega_1)\frac{dW}{ d\omega_1}\,.
\end{eqnarray}

\begin{figure*}[t]
\includegraphics[scale=0.5]{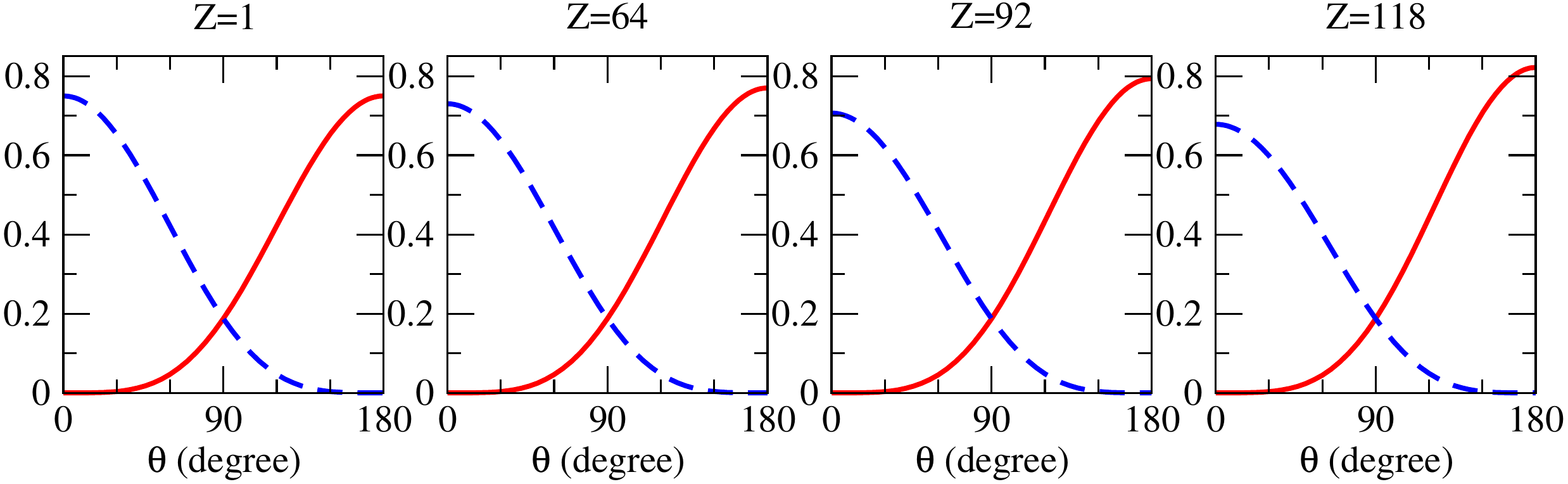}
\caption{The normalized differential transition probabilities $\left[\left(\frac{dW^{(e)(\lambda_1 \lambda_2)}}{\sin\theta d\theta d\omega_1}\right) \left(\frac{dW^{(e)}}{d\omega_1}\right)^{-1}\right]$ as a function of the angle between the momenta of the emitted photons ($\theta$) for $x=1/2$.
The results for the photon linear polarization $\zhep^{(45^{\circ})}$, $\zhep^{(135^{\circ})}$ and the circular polarization considered in Subsections \ref{lp2} and \ref{circ} are presented, respectively.
In the case of the photon linear polarization $\zhep^{(45^{\circ})}$, $\zhep^{(135^{\circ})}$, the blue dashed line represents angular dependence of the differential transition probability for the photons with equal polarizations
[see
\Eq{38}],
the red solid line gives angular dependence of the differential transition probability for the photons with different polarizations
[see
\Eq{39}]. 
In the case of the circular photon polarization,  the blue dashed line represents angular dependence of the differential transition probability for the photons with different polarizations[see
\Eq{43}
],
the red solid line gives angular dependence of the differential transition probability for the photons with equal polarizations
[see
\Eq{42}].
The data are presented for $Z=1,  64,  92,  118$.
}
\label{fig4}
\end{figure*}

The numerical results for the differential transition probabilities (for $x=1/2$) as a function of the angle $\theta$ are presented in
\Fig{fig4}.
The blue dashed line shows the differential transition probabilities for the emission of photons with the same polarizations ($\lambda_1=\lambda_2$).
The red solid line gives the differential transition probabilities for the emission of photons with different polarizations ($\lambda_1\ne \lambda_2$).
The transition probabilities for the emitted photons with this linear polarization are explicitly related to the transition probabilities for the circularly polarized photons, which is discussed below.

\subsubsection{Circular polarization}
\label{circ}
The polarization vectors of the emitted photons with the circular polarization read as
\begin{eqnarray}
\zhep^{(+)}_i(\theta_i,\varphi_i)
&=&
\frac{1}{\sqrt{2}}(\zhe_i^{(1)}(\theta_i,\varphi_i)+i\zhe_i^{(2)}(\theta_i,\varphi_i))\,,
\\
\zhep^{(-)}_i(\theta_i,\varphi_i)
&=&
\frac{1}{\sqrt{2}}(\zhe_i^{(1)}(\theta_i,\varphi_i)-i\zhe_i^{(2)}(\theta_i,\varphi_i))\,.
\end{eqnarray}

Using \Eq{eqn210122n07}, we can write the differential transition probabilities in the two-parameters approximation as 
\begin{eqnarray}
\frac{dW^{(++)}}{\sin\theta d\theta d\omega_1}
&=& \label{42}
\frac{dW^{(--)}}{\sin\theta d\theta d\omega_1}
=
\frac{3}{4}\sin^4\frac{\theta}{2} \fun(\theta,\omega_1)\frac{dW}{ d\omega_1}\,,
\\
\frac{dW^{(+-)}}{\sin\theta d\theta d\omega_1}
&=& \label{43}
\frac{dW^{(-+)}}{\sin\theta d\theta d\omega_1}
=
\frac{3}{4}\cos^4\frac{\theta}{2} \fun(\theta,\omega_1)\frac{dW}{ d\omega_1}\,.
\end{eqnarray}

The differential transition probability as a function of the angle $\theta$ is presented in
\Fig{fig4}.
The results for circular polarization coincide exactly the opposite with the results for linear polarization $\zhep^{(45^{\circ})}$, $\zhep^{(135^{\circ})}$.
The red solid line gives the differential transition probabilities for emission of photons with the different polarizations ($\lambda_1=\lambda_2$).
The blue dashed line shows the differential transition probabilities for the emission of photons with the same polarizations ($\lambda_1\ne \lambda_2$).

\subsection{Contribution of the negative continuum}

According to Eqs.~\Br{u1}, \Br{ub}, both the positive and negative energy parts of the Dirac spectrum contribute to the two-photon transition probabilities.
The contribution of the negative energy part was investigated in the work of Labzowsky \cite{labzowsky2005jpb38-265}, where it was shown that its contribution to the total transition probabilities is small.
The contribution of the negative continuum to the differential transition probabilities was studied in 
\cite{surzhykov2009negcontinuum}.
It was noticed that contribution of the negative continuum to M1M1 and E2E2 multipole of the two-photon transitions is of great importance \cite{surzhykov2009negcontinuum}. 
Since the asymmetry of the differential transition probability is a consequence of the interference of the E1E1 multipoles with M1M1 and E2E2 multipoles, the contribution of the negative continuum to the asymmetry factor is very large.
In Tables
\ref{tab5} and \ref{tab-muon-posneg}
we present the results of the calculations of the  transition probabilities, where we give separately the contributions of the positive and negative energy intermediate states for one-electron and one-muon ions, respectively.
The calculation was performed in the transverse gauge (see \Eq{Atr}).
The data in Table \ref{tab5} show that the negative continuum gives the dominant contribution to the asymmetry even for light ions,
while the positive energy intermediate states gives the main contribution to the transition probability. 
In Table \ref{tab-muon-posneg}
we can see that the contribution of the negative continuum for one-muon ions is very small even for heavy ions.

\subsection{Nonresonant corrections}
\label{Nonresonant_corrections}
The exited energy level is usually characterized by two parameters: the energy and the width of the level. 
This is the so-called resonant approximation \cite{andreev08pr}.
In this approximation, the line profile is described by the Lorentz contour, and the energy and width of the level do not depend on the particular process of measurement.
If we go beyond the resonant approximation, nonresonant corrections arise, which lead to asymmetry of the line profile \cite{low52}.
Nonresonant corrections are usually very small, but they are of importance since they indicate the limit to which the concept of energy for an excited atomic state has a physical meaning \cite{andreev08pr}.
This corrections were investigated in many works
\cite{labzowsky01, jentschura02, labzowsky04p3271,labzowsky07}. 
In some precision experiments, nonresonant corrections are already taken into account when determining the accuracy of the experiment
\cite{PhysRevLett.119.263002}.
These corrections should also be important for precision measurements with muon ions
\cite{pohl2010}.
We would like to note that, for the light ions, the asymmetry factors for electron and muon ions are of the same order of magnitude.

Since the asymmetry factor has a nonzero value even for light ions, it can lead to nonresonant corrections to the energy levels.
In particular, for the hydrogen atom, the asymmetry factor $a(x)$ (for $x=1/2$) is equal to $6.1\times 10^{-6}$. 
This asymmetry factor should be compared with the declared accuracy of precise measurements of frequency of $1s-2s$ two-photon transition in atomic hydrogen \cite{PhysRevLett.110.230801}, which is $4.5\times 10^{-15}$.
This measurement was performed in an experiment in which the reverse process was studied: two-photon excitation of $1s$ state.
The asymmetry factor gives a correction to the two-photon transition probabilities due to the nonzero angle between the photon momenta.
If the angle spread between momenta equals to $\delta\theta$, then this should lead to a relative uncertainty for the  transition probability (see \Eq{eqn210121n02})
\begin{widetext}
\begin{eqnarray}
\left(\frac{dW}{\sin{\theta} d\theta d\omega_1}(\delta\theta)-\frac{dW}{\sin{\theta}d\theta d\omega_1}(0)\right)\left(\frac{dW}{\sin{\theta}d\theta d\omega_1}(0)\right)^{-1}&=&a(x)(1-\cos\delta\theta)\,.
\label{44}
\end{eqnarray}
\end{widetext}
If $\delta\theta$ is about $1^{\circ}$, then the relative uncertainty given by \Eq{44} is about $9.3\times 10^{-10}$.
The relative difference between the differential transition probabilities for $0^{\circ}$ and $180^{\circ}$ (given by \Eq{44} with $\delta\theta=180^{\circ}$) is $1.2\times 10^{-5}$.
We note that in the experiment \cite{PhysRevLett.110.230801} a set of mirrors was used. 
Thus, photons were absorbed at both $0^{\circ}$ and $180^{\circ}$ angles.
For absorption of photons with $0^{\circ}$ angle between the momenta, the asymmetry factor decreases the transition probability, while for absorption with $180^{\circ}$, the asymmetry factor increases the transition probability. 
Accordingly, the presence of absorption of photons at different angles should significantly reduce the described  uncertainty. 
Nevertheless, in principle, the asymmetry factor should be taken into account as a source of nonresonant corrections.

\section{Summary}
We investigated the radiative decay of $2s$ state of one-electron and one-muon ions with respect to the polarization of the emitted photons.
The investigation was performed for the ions with nuclear charge numbers $1\le Z \le 120$.
The particular attention was paid to the role of the two-photon decay channel.
For both electron and muon ions the most long-lived state of the L-shell is the $2s$ state.
The radiative decay of the $2s$ state in the electron and muon ions is qualitatively different.
In particular, in contrast to electron ions, in the case of muon ions the cascade ($2s\to2p_{3/2}\to1s$ and $2s\to2p_{1/2}\to1s$) channels are of great importance.
For the muon ions taking into account the nuclear size corrections may change the transition probability by several orders of magnitude.

The two-parameter approximation is introduced, which makes it possible to describe with high accuracy the two-photon angular-differential transition probability for the polarized emitted photons.
The accuracy of this approximation is $10^{-3}\%$ for light ions, remaining within $1\%$ even for the super heavy ions (for the photons with equal energies).
The parameters of the approximation are the total (or energy-differential) transition probability and the asymmetry factor, they are listed in the tables.
Within the two-parameter approximation, the asymmetry factor completely determines the asymmetry of the differential transition probability.
For the one-muon ions the asymmetry is very small.
For the one-electron ions the main contribution to the  asymmetry factor is made by the negative continuum of the Dirac spectrum.
Using the two-parameter approximation, we investigated the various polarizations of the emitted photons.
The angular dependence of the differential transition probabilities for the emission of circularly polarized photons is clearly related to the transition probabilities for linearly polarized photons.
A nonzero asymmetry factor even for light ions can be a source of nonresonant corrections, which can be important for precision experiments.


\begin{table*}
 \caption{\label{tab-el-2p2p2s}
The transition probabilities ($W^{(e)}$, in s$^{-1}$)  for one-electron ions.
The digits in square brackets refer to the power of 10.
The first column gives the charge of the atomic nuclei ($Z$).
The next four columns present the one-photon (E1) transition probabilities for the $2p_{1/2}\to 1s$
and $2p_{3/2}\to 1s$ transitions, respectively.
The next two columns present the sum of one-photon (M1) and two-photon transition probabilities for the $2s\to 1s$ transition.
The last column gives the  $2s\to 2p_{1/2}$  transitions probabilities.
The values $W^{(e)}_0$ are calculated with the point-like nucleus.
The values $W^{(e)}$ are calculated with the Fermi distribution of the nuclear charge density.
} 

\begin{tabular}{r|l|l|l|l|l|l|l}
\hline
\multicolumn{1}{c|}{Nucleus}
&\multicolumn{2}{c|}{$2p_{1/2}\to 1s$}
&\multicolumn{2}{c|}{$2p_{3/2}\to 1s$}
&\multicolumn{2}{c|}{$2s\to 1s$ ($1\gamma+2\gamma$)}
&\multicolumn{1}{c}{$2s\to 2p_{1/2}$}\\
\hline
\multicolumn{1}{c|}{Z}
&\multicolumn{1}{c|}{$W^{(e)}_0$}
&\multicolumn{1}{c|}{$W^{(e)}$}
&\multicolumn{1}{c|}{$W^{(e)}_0$}
&\multicolumn{1}{c|}{$W^{(e)}$}
&\multicolumn{1}{c|}{$W^{(e)}_0$}
&\multicolumn{1}{c|}{$W^{(e)}$}
&\multicolumn{1}{c}{$W^{(e)}$}
\\
\hline
1
&6.26835[8]&6.26835[8]
&6.26824[8]&6.26824[8]
&8.22906&8.22906
&8.56912[-21]\\

10
&6.27225[12]&6.27225[12]
&6.26060[12]&6.26060[12]
&8.22575[6]&8.22574[6]
&4.26634[-8]\\

50
&3.98005[15]&3.97985[15]
&3.79354[15]&3.79338[15]
&4.01596[11]&4.01592[11]
&6.34525[1]\\

92
&4.72601[16]&4.72033[16]
&3.95022[16]&3.94939[16]
&1.96240[14]&1.96244[14]
&2.63046[7]\\

120
&1.37847[17]&1.36493[17]
&9.66319[16]&9.77117[16]
&4.74519[15]&4.74417[15]
&1.49830[11]\\
\hline
\end{tabular}
\end{table*}

\begin{table*}
\caption{\label{tab-mu-2p2p2s}
The transition probabilities ($W^{(\mu)}$, in s$^{-1}$)  for one-muon ions.
The first column gives the charge of the atomic nuclei ($Z$).
The next four columns present the one-photon (E1) transition probabilities for the $2p_{1/2}\to 1s$
and $2p_{3/2}\to 1s$ transitions, respectively.
The last two columns present the sum of one-photon (M1) and two-photon transition probabilities for the $2s\to 1s$ transition.
The values $W^{(\mu)}_0$ are calculated with the point-like nucleus.
The values $W^{(\mu)}$ are calculated with the Fermi distribution of the nuclear charge density (the nuclear recoil and the vacuum polarization corrections are also taken into account).
} 
\begin{tabular}{r|l|l|l|l|l|l}
\hline
\multicolumn{1}{c|}{Nucleus}
&\multicolumn{2}{c|}{$2p_{1/2}\to 1s$}
&\multicolumn{2}{c|}{$2p_{3/2}\to 1s$}
&\multicolumn{2}{c}{$2s\to 1s$ ($1\gamma+2\gamma$)}\\
\hline
\multicolumn{1}{c|}{Z}
&\multicolumn{1}{c|}{$W^{(\mu)}_0$}
&\multicolumn{1}{c|}{$W^{(\mu)}$}
&\multicolumn{1}{c|}{$W^{(\mu)}_0$}
&\multicolumn{1}{c|}{$W^{(\mu)}$}
&\multicolumn{1}{c|}{$W^{(\mu)}_0$}
&\multicolumn{1}{c}{$W^{(\mu)}$}
\\
\hline
1
&1.29610[11]&1.16600[11]
&1.29607[11]&1.16598[11]
&1.70151[3]&1.53071[3]\\

2
&2.07379[12]&2.02137[12]
&2.07364[12]&2.02122[12]
&1.08886[5]&1.06175[5]\\

5
&8.10183[13]&8.04560[13]
&8.09807[13]&8.04186[13]
&2.65686[7]&2.64501[7]\\

&&6[13]$^{a}$
&&
&& \\

10
&1.29690[15]&1.28259[15]
&1.29449[15]&1.28030[15]
&1.70082[9] &2.13163[[9]\\

&&1[15]$^{a}$
&&
&& \\

20
&2.07896[16]&1.96523[16]
&2.06350[16]&1.95388[16]
&1.12827[11]&8.88038[11]\\

30
&1.05580[17]&9.09875[16]
&1.03810[17]&9.02417[16]
&1.52467[12]&4.57589[13]\\

&&1[17]$^{a}$
&&
&& \\

40
&3.35164[17]&2.52880[17]
&3.25149[17]&2.51522[17]
&1.25584[12]&5.93759[14]\\

50
&8.22949[17]&5.18286[17]
&7.84383[17]&5.20931[17]
&8.30638[13]&3.81497[15] \\

60
&1.71835[18]&8.76268[17]
&1.60181[18]&8.95361[17]
&4.59891[14]&1.48179[16]\\

70
&3.20931[18]&1.26516[18]
&2.91112[18]&1.31995[18]
&2.15766[15]&4.32850[16]\\

80
&5.52446[18]&1.71216[18]
&4.84830[18]&1.82341[18]
&8.78912[15]&9.53850[16]\\

90
&8.93288[18]&2.10984[18]
&7.53338[18]&2.29237[18]
&3.20313[16]&1.83676[17]\\

92
&9.77188[18]&2.12891[18]
&8.16780[18]&2.32133[18]
&4.10465[16]&2.09096[17] \\

100
&1.37345[19]&2.51872[18]
&1.10367[19]&2.78650[18]
&1.07708[17]&3.09386[17]\\

110
&2.02189[19]&2.83524[18]
&1.53073[19]&3.18497[18]
&3.45306[17]&4.79458[17]\\

118
&2.67118[19]&3.07752[18]
&1.90552[19]&3.49419[18]
&8.71521[17]&6.44089[17]\\

120
&2.84985[19]&3.13174[18]
&1.99841[19]&3.56431[18]
&1.10201[18]&6.89189[17]\\
\hline
\multicolumn{3}{l}{$\phantom{qqqqq}^{a}$ \cite{MISSIMER1985179}}\\
\end{tabular}
\end{table*}

\begin{table*}
\caption{\label{tab21}
The transition probabilities ($W^{(e)}$, in s$^{-1}$) for
one- and two-photon $2s\to1s$ transitions in one-electron ions.
The value of $p_W$ shows the power dependence on $Z$ of the corresponding transition probability ($W^{(e)}\sim Z^{p_W}$).
In the first column the atomic number of the ion ($Z$)  is indicated.
In the next two columns the one-photon transition probabilities and their power dependence on $Z$ are given.
In the columns marked `E1E1: $2s\to1s$' we give the two-photon transition probabilities with emission of E1E1 photons ($W^{(e)}$, in s$^{-1}$) and the corresponding results of work \cite{labzowsky2005jpb38-265} ($W^{(e)a}$) together with their power dependence on $Z$.
The multicolumn `$2s\to1s$, Total 2$\gamma$' presents the results of exact calculation of transition probabilities: $W^{(e)}$ -- the total transition probability, non-spin-flip and spin-flip -- transition probabilities in which the initial state does not change or changes the projection of the total angular momentum, respectively.
We note that the spin-flip for one-photon M1 transition is 2/3 of the total transition probability $W^{(e)}$, while the non-spin-flip is 1/3 $W^{(e)}$.
} 

\begin{tabular}{r|l|l|l|l|l|l|l|l|l|l|l}
\hline
\multicolumn{1}{c|}{Nucleus}
&\multicolumn{2}{c|}{M1: $2s\to 1s$}
&\multicolumn{3}{c|}{E1E1: $2s\to 1s$}
&\multicolumn{6}{c}{$2s\to 1s$, Total 2$\gamma$}\\
\hline
\multicolumn{1}{c|}{Z}
&\multicolumn{1}{c|}{$W^{(e)}$}
&\multicolumn{1}{c|}{$p_W$}
&\multicolumn{1}{c|}{$W^{(e)}$}
&\multicolumn{1}{c|}{$W^{(e)a}$}
&\multicolumn{1}{c|}{$p_W$}
&\multicolumn{1}{c|}{$W^{(e)}$}
&\multicolumn{1}{c|}{$p_W$}
&\multicolumn{1}{c|}{non-spin-flip}
&\multicolumn{1}{c|}{$p_{\txt{nsf}}$}
&\multicolumn{1}{c|}{spin-flip}
&\multicolumn{1}{c}{$p_{\txt{sf}}$}
\\
\hline
1 & 2.49592[-6]& 10.00  &   8.22906  &$8.22906^a$ &6.00&   8.22906  & 6.00& 8.22906 &6.00&3.88291[-9] &9.99\\

10& 2.51003[4] &10.01 &    8.20064[6]&$8.1923[6]^a$&5.99 &    8.20065[6] &5.99 & 8.20061[6]&5.99&3.15349[1]&9.64\\

20& 2.61488[7]& 10.05 & 5.19513[8]&$5.1901[8]^a$&5.97 &5.19515[8] &5.97 &5.19492[8]&5.97&2.30865[4] &9.43\\

30& 1.55241[9]&10.11  & 5.82109[9]&$5.8151[9]^a$& 5.94&5.82125[9]&5.94 & 5.82019[9]& 5.94&1.05200[6]&9.36\\

40& 2.87414[10]&10.20 & 3.19862[10]&$3.1954[10]^a$& 5.90 &3.19889[10]&  5.90&3.19735[10]& 5.90 &1.54080[7]&9.28\\

50& 2.82905[11]& 10.32&1.18662[11] &$1.1854[11]^a$&5.84  &1.18686[11]& 5.84 &1.18565[11]&5.84&1.21404[8]&9.21 \\

60& 1.87950[12]&10.48 &3.42645[11] & $3.4229[11]^a$&5.78& 3.42797[11]&  5.78 & 3.42150[11]&5.78&6.47328[8]& 9.15\\

64&3.70310[12]&10.56&4.97148[11]&&5.75&4.97436[11]&5.75
&4.96269[11]&5.74&1.16734[9]&9.11\\

70& 9.58288[12]&10.69&8.30599[11]&$8.2975[11]^a$&5.70  &8.31297[11] &5.70 & 8.28657[11]&5.69&2.63989[9]& 9.09\\

80& 4.05532[13]&10.96&1.76726[12]&$1.7655[12]^a$&5.59  &1.76988[12] & 5.60&1.76102[12] &5.58&8.85741[9]& 9.04\\

90& 1.50037[14]& 11.35&3.39348[12]&$3.3899[12]^a$&5.46  &3.40186[12]&5.47 & 3.37619[12]&5.44&2.56687[10]&9.03\\

92& 1.92408[14]&11.44&3.82557[12]&$3.8216[12]^a$& 5.43 &3.83600[12] &5.44 &3.80469[12]&5.41&3.13168[10]&9.00 \\

100& 5.04074[14]& 11.79& 5.98484[12] &$5.9782[12]^a$ &5.28  &6.00879[12]&  5.30& 5.94218[12]& 5.25&6.66209[10]&9.15\\

110& 1.58119[15]& 12.45& 9.80101[12]&&5.04  &9.86357[12] &5.07 &9.70146[12]&4.99&1.62121[11] &9.87\\

118& 3.81066[15]& 12.81& 1.38978[13]&&5.02  &1.40264[13] &5.02&1.36779[13]& 4.80&3.48470[11]&13.40\\

120&4.72890[15] &12.94 &1.51115[13]&& 5.06 &1.52650[13] &5.12&1.48250[13]&4.80&4.39623[11]&15.44\\
\hline
\multicolumn{6}{l}{$\phantom{qqqqq}^{a}$ \cite{labzowsky2005jpb38-265}}\\
\end{tabular}
\end{table*}

\begin{table*}
 \caption{
\label{tab-tp-2s-muon}
The transition probabilities ($W^{(\mu)}$, in s$^{-1}$) for
the one- and two-photon decay of the $2s$ state of one-muon ions.
The values $W^{(\mu)}_0$ are calculated with the point-like nucleus.
The values $W^{(\mu)}$ are calculated with the Fermi
distribution of the nuclear charge density (the nuclear recoil and the vacuum polarization corrections are also taken into account).
The notations are the same as in
Table~\ref{tab-mu-2p2p2s}.
} 
\begin{tabular}{r|l|l|l|l|l|l}
\hline
\multicolumn{1}{c|}{Nucleus}
&\multicolumn{2}{c|}{M1: $2s\to 1s$}
&\multicolumn{1}{c|}{E1: $2s\to2p_{1/2}$}
&\multicolumn{1}{c|}{E1: $2s\to2p_{3/2}$}
&\multicolumn{2}{c}{$2s\to 1s$, Total 2$\gamma$}\\
\hline
\multicolumn{1}{c|}{Z}
&\multicolumn{1}{c|}{$W^{(\mu)}_0$}
&\multicolumn{1}{c|}{$W^{(\mu)}$}
&\multicolumn{1}{c|}{$W^{(\mu)}$}
&\multicolumn{1}{c|}{$W^{(\mu)}$}
&\multicolumn{1}{c|}{$W^{(\mu)}_0$}
&\multicolumn{1}{c}{$W^{(\mu)}$}
\\
\hline
1
&5.16078[-4]&4.66233[-4]
&2.25265
&5.09542
&1.70151[3]&1.53071[3]\\

2
&5.28554[-1]&5.18875[-1]
&1.53770[2]
&4.16282[2]
&1.08885[5]&1.06174[5]\\

5
&5.04671[3]&4.99412[3]
&4.66810[3]
&1.88840[2]
&2.65636[7]&2.64451[7]\\

&&5[3]$^{a}$&1[4]$^{a}$&&&3[7]$^{a}$\\

10
&5.18997[6]&4.72237[6]&2.17920[8]&2.20546[8]&1.69563[9]&2.12691[9]\\

&&5[6]$^{a}$&1[9]$^{a}$&&&2[9]$^{a}$\\

20
&5.40686[9]&3.44581[9]
&3.62043[11]
&4.17798[11]
&1.07420[11]&8.84592[11]\\

30
&3.21013[11]&1.17226[11]
&2.02420[13]
&2.42670[13]
&1.20366[12]&4.56417[13]\\

&&1[11]$^{a}$&5[13]$^{a}$&&&4[11]$^{a}$\\

40
&5.94395[12]&1.14696[12]
&2.65386[14]
&3.21458[14]
&6.61441[12]&5.92612[14]\\

50
&5.85222[13]&5.56180[12]&1.69154[15]&2.09944[15]&2.45416[13]&3.80941[15] \\

60
&3.89006[14]&1.82667[13]
&6.50114[15]
&8.25557[15]
&7.08851[13]&1.47996[16]\\

70
&1.98575[15]&4.33285[13]
&1.85809[16]
&2.45839[16]
&1.71910[14]&4.32417[16]\\

80
&8.42309[15]&9.37768[13]
&4.05345[16]
&5.46345[16]
&3.66026[14]&9.52912[16]\\

90
&3.13278[16]&1.66535[14]
&7.67106[16]
&1.06633[17]
&7.03501[14]&1.83509[17]\\

92
&4.02531[16]&1.72935[14]&8.64745[16]&1.22282[17]&7.93361[14]&2.08923[17] \\

100
&1.06466[17]&2.80172[14]
&1.27813[17]
&1.81081[17]
&1.24176[15]&3.09106[17]\\

110
&3.43275[17]&4.13532[14]
&1.95286[17]
&2.83513[17]
&2.03054[15]&4.79044[17]\\

118
&8.68675[17]&5.49249[14]
&2.60176[17]
&3.83097[17]
&2.84567[15]&6.43540[17]\\

120
&1.09894[18]&5.85407[14]&2.77829[17]&4.10504[17]&3.06912[15]&6.88604[17]\\
\hline
\multicolumn{3}{l}{$\phantom{qqqqq}^{a}$ \cite{MISSIMER1985179}}\\
\end{tabular}
\end{table*}

\begin{table*}
 \caption{\label{tab-tp-2s-spinf-muon}
The transition probabilities ($W^{(\mu)}$, in s$^{-1}$) for
one- and two-photon decay of the $2s$ state of one-muon ions.
The notations are the same as in
Table~\ref{tab-tp-2s-muon}.
In the second column, the energy difference between $2s$ and $1s$ muon states is given
($\Delta E=\dee_{2s}-\dee_{1s}$, in keV)
} 
\begin{tabular}{r|l|l|l|l|l}
\hline
\multicolumn{1}{c|}{Nucleus}
&\multicolumn{1}{c|}{Frequency}
&\multicolumn{1}{c|}{E1E1: $2s\to1s$}
&\multicolumn{3}{c}{$2s\to1s$, Total 2$\gamma$}\\
\hline
\multicolumn{1}{c|}{Z}
&\multicolumn{1}{c|}{(keV)}
&\multicolumn{1}{c|}{$W^{(\mu)}$}
&\multicolumn{1}{c|}{$W^{(\mu)}$}
&\multicolumn{1}{c|}{non-spin-flip}
&\multicolumn{1}{c}{spin-flip}
\\
\hline
1
&1.89818&1.53071[3]
&1.53071[3]&1.53071[3]&6.43978[-7]\\

2
&8.22384&1.06174[5]
&1.06174[5]&1.06174[5]&6.50581[-4]\\

5
&5.22860[1]&2.64451[7]
&2.64451[7]&2.64428[7]&2.35653[3]\\

10
&2.07693[2]&2.12691[9]
&2.12691[9]&1.96354[9]&1.63372[8]\\

20
&7.91683[2]&8.84588[11]
&8.84592[11]&5.99008[11]&2.85584[11]\\

30
&1.64027[3]&4.56406[13]
&4.56417[13]&2.94113[13]&1.62303[13]\\

40
&2.64523[3]&5.92577[14]
&5.92612[14]&3.78631[14]&2.13981[14]\\

50
&3.70203[3]&3.80899[15]
&3.80941[15]&2.43066[15]&1.37875[15]\\

60
&4.77824[3]&1.47970[16]
&1.47996[16]&9.44594[15]&5.35368[15]\\

70
&5.77201[3]&4.32310[16]
&4.32417[16]&2.76732[16]&1.55685[16]\\

80
&6.82037[3]&9.52598[16]
&9.52912[16]&6.10387[16]&3.42524[16]\\

90
&7.74204[3]&1.83435[17]
&1.83509[17]&1.17823[17]&6.89703[16]\\

92
&7.82436[3]&2.08837[17]
&2.08923[17]&1.34355[17]&7.45678[16]\\

100
&8.67626[3]&3.08954[17]
&3.09106[17]&1.98734[17]&1.10372[17]\\

110
&9.46714[3]&4.78775[17]
&4.79044[17]&3.08634[17]&1.70410[17]\\

118
&1.00866[4]&6.43142[17]
&6.43540[17]&4.15117[17]&2.28423[17]\\

120
&1.02323[4]&6.88168[17]
&6.88604[17]&4.44322[17]&2.44282[17]\\
\hline
\end{tabular}

\end{table*}

\begin{table*}
 \caption{\label{tab-el-rad}
The bound energies and the root mean square radii for electron ions.
In the first two columns, the nuclear charge ($Z$) and the nuclear root mean square charge radii ($R$, in fm) are given.
In the next columns, the bound energies ($E^{(e)}=\dee^{(e)}-m_e c^2$, in keV) and the root mean square radii
($\langle\psi|r^2|\psi\rangle^{1/2}$, in fm) are presented for the corresponding electron states.
} 

\begin{tabular}{r|l|l|l|l|l|l|l|l|l}
\hline
\multicolumn{2}{c|}{Nucleus}
&\multicolumn{2}{c|}{$1s$}
&\multicolumn{2}{c|}{$2s$}
&\multicolumn{2}{c|}{$2p_{1/2}$}
&\multicolumn{2}{c}{$2p_{3/2}$}\\
\hline
\multicolumn{1}{c|}{Z}
&\multicolumn{1}{c|}{R}
&\multicolumn{1}{c|}{$E^{(e)}$}
&\multicolumn{1}{c|}{$r^{(e)}$}
&\multicolumn{1}{c|}{$E^{(e)}$}
&\multicolumn{1}{c|}{$r^{(e)}$}
&\multicolumn{1}{c|}{$E^{(e)}$}
&\multicolumn{1}{c|}{$r^{(e)}$}
&\multicolumn{1}{c|}{$E^{(e)}$}
&\multicolumn{1}{c}{$r^{(e)}$}
\\
\hline
1 &0.8791
&-1.360587283[-2]&91654.8 
&-3.401479529[-3]&342940. 
&-3.401479530[-3]&289836. 
&-3.401434246[-3]&289841. 
\\

10&3.0053
&-1.36238&9151.37 
&-0.34071&34233.1 
&-0.34071&28923.1 
&-0.34026&28970.1
\\

50&4.6266
&-3.52266[1]&1759.47
&-8.88410&6543.62
&-8.88437&5483.07
&-8.57551&5725.50
\\

92&5.860
&-1.32081[2]&846.916
&-3.41777[1]&3101.99
&-3.42111[1]&2525.50
&-2.96498[1]&3016.38
\\

120&6.330
&-2.59627[2]&543.913
&-6.97852[1]&1960.51
&-7.06350[1]&1500.79
&-5.15841[1]&2236.49
\\

\hline
\end{tabular}
\end{table*}

\begin{table*}
\caption{\label{tab-muon-rad}
The bound energies and the root mean square radii for muon ions.
In the first two columns, the nuclear charge ($Z$) and the nuclear root mean square charge radii ($R$, in fm) are given.
In the next columns, the bound energies ($E^{(\mu)}=\dee^{(\mu)}-m_{\mu} c^2$, in keV) and the root mean square radii
($\langle\psi|r^2|\psi\rangle^{1/2}$, in fm) are presented for the corresponding muon states.
} 

\begin{tabular}{r|l|l|l|l|l|l|l|l|l}
\hline
\multicolumn{2}{c|}{Nucleus}
&\multicolumn{2}{c|}{$1s$}
&\multicolumn{2}{c|}{$2s$}
&\multicolumn{2}{c|}{$2p_{1/2}$}
&\multicolumn{2}{c}{$2p_{3/2}$}\\
\hline
\multicolumn{1}{c|}{Z}
&\multicolumn{1}{c|}{R}
&\multicolumn{1}{c|}{$E^{(\mu)}$}
&\multicolumn{1}{c|}{$r^{(\mu)}$}
&\multicolumn{1}{c|}{$E^{(\mu)}$}
&\multicolumn{1}{c|}{$r^{(\mu)}$}
&\multicolumn{1}{c|}{$E^{(\mu)}$}
&\multicolumn{1}{c|}{$r^{(\mu)}$}
&\multicolumn{1}{c|}{$E^{(\mu)}$}
&\multicolumn{1}{c}{$r^{(\mu)}$}
\\
\hline
1 &0.8791
&-2.53057&492.842 
&-6.32394[-1]& 1844.61
&-6.32192[-1]& 1559.44
&-6.32184[-1]& 1559.46
\\

10&3.0053
&-2.77410[2]&44.9618 
&-6.97169[1]&167.328
&-7.01762[1]&140.450 
&-7.00826[1]&140.677 
\\

50&4.6266
&-5.23928[3]& 11.6014
&-1.53726[3]& 37.7515
&-1.81440[3]& 27.0449
&-1.76858[3]& 27.8725
\\

92&5.860
&-1.21496[4]&8.88120 
&-4.32520[3]&24.4212
&-5.93616[3]&15.4070 
&-5.70775[3]&16.0846 
\\

120&6.330
&-1.68862[4]& 8.16188
&-6.65385[3]& 20.5221
&-9.46687[3]& 12.7713
&-9.10565[3]& 13.3167
\\

\hline
\end{tabular}
\end{table*}

\begin{table*}
\caption{\label{tab-muon-tptp-details}
Corrections to the two-photon transition probabilities for one-muon ions
(in s${}^{-1}$).
In the first column the nuclear charge ($Z$) is given.
The multicolumn labeled 'point' presents the results of calculation for the point-like nucleus: the transition probability ($W^{(\mu)}_0$),
its power dependence on $Z$ ($W^{(\mu)}_0\propto Z^{p_{W_0}}$), the asymmetry parameter $A_0$ and its power dependence on $Z$ ($A_0\propto Z^{p_{A_0}}$).
The numbers in parentheses indicate the accuracy of the two-parameter approximation defined by
Eqs.~\Br{eqn210122n01}-\Br{A}.
The column labeled 'Fermi' gives the results of calculation with the Fermi distribution of the nuclear charge density.
The column labeled 'Fermi, NR' gives the results of calculation with the Fermi distribution of the nuclear charge density and the nuclear recoil correction taken into account.
The column labeled 'Fermi, NR, VP' gives the results of calculation with the Fermi distribution of the nuclear charge density, the nuclear recoil correction and the electron vacuum polarization (in the Uehling approximation) corrections taken into account.
}
\begin{tabular}{r|ll|ll|ll|ll|ll}
\hline
\multicolumn{1}{c|}{}
&\multicolumn{4}{c|}{point}
&\multicolumn{2}{c|}{Fermi}
&\multicolumn{2}{c|}{Fermi, NR}
&\multicolumn{2}{c}{Fermi, NR, VP}\\
\hline
\multicolumn{1}{c|}{Z}
&\multicolumn{1}{c}{$W_0^{(\mu)}$}
&\multicolumn{1}{c|}{$p_{W_0}$}
&\multicolumn{1}{c}{$\Asym_0$}
&\multicolumn{1}{c|}{$p_{\Asym_{0}}$}
&\multicolumn{1}{c}{$W^{(\mu)}$}
&\multicolumn{1}{c|}{$\Asym$}
&\multicolumn{1}{c}{$W^{(\mu)}$}
&\multicolumn{1}{c|}{$\Asym$}
&\multicolumn{1}{c}{$W^{(\mu)}$}
&\multicolumn{1}{c}{$\Asym$}
\\
\hline
1
&1.70151[3]&6.00&-2.48681(4)[-5]&-0.17
&1.70149[3]&-2.48683(4)[-5]
&1.52939[3]&-2.4868(1)[-5]
&1.53071[3]&-2.4861(1)[-5]\\

10
&1.69563[9]&5.99&4.2702(1)[-4]&2.01
&2.38340[9]&2.9(2)[-4]
&2.34722[9]&2.9(2)[-4]
&2.12691[9]&3.3(1)[-4]\\

50
&2.45416[13]&5.84&1.140(2)[-2]&2.13
&3.79574[15]&2.9(8)[-5]
&3.78207[15]&2.9(8)[-5]
&3.80941[15]&3.0(8)[-5]\\

92
&7.93239[14]&5.44&4.3(1)[-2]&2.17
&2.06754[17]&2.1(6)[-5]
&2.06579[17]&1.9(6)[-5]
&2.08923[17]&1.9(6)[-5]\\

120
&3.06912[15]&3.38&6.8(2)[-2]&0.27
&6.82392[17]&1.9(6)[-5]
&6.82120[17]&2.0(6)[-5]
&6.88604[17]&2.0(6)[-5]\\
\hline
\end{tabular}
\end{table*}

\begin{table*}
\caption{\label{tab-muon-tptp-ball}
Corrections to the two-photon transition probabilities for one-muon ions
(in s${}^{-1}$).
The column labeled 'Fermi' gives the results of calculation with the Fermi distribution of the nuclear charge density.
The column labeled 'sphere' shows the results of calculations with the nucleus considered as a homogeneously charged sphere. 
}
\begin{tabular}{r|l|l|l|l}
\hline
\multicolumn{1}{c|}{}
&\multicolumn{2}{c|}{Fermi}
&\multicolumn{2}{c}{sphere}\\
\hline
\multicolumn{1}{c|}{Z}
&\multicolumn{1}{c}{$W^{(\mu)}$}
&\multicolumn{1}{c|}{$\Asym$}
&\multicolumn{1}{c}{$W^{(\mu)}$}
&\multicolumn{1}{c}{$\Asym$}
\\
\hline

10
&2.12691[9]&3.3(1)[-4]
&2.13988[9]&3.3(2)[-4]\\

50
&3.80941[15]&3.0(8)[-5]
&3.98007[15]&2.8(1)[-5]\\

92
&2.08923[17]&1.9(6)[-5]
&2.16710[17]&1.9(6)[-5]\\

120
&6.88604[17]&2.0(6)[-5]
&7.11310[17]&1.9(6)[-5]\\
\hline
\end{tabular}
\end{table*}

\begin{table}
\caption{\label{tab1}
The transition probabilities ($W^{(e)}$, in s$^{-1}$)  for two-photon decay of $2s$-electron state and 
the asymmetry factor ($\Asym$).
The numbers in parentheses indicate the accuracy of the two-parameter approximation defined by
Eqs.~\Br{eqn210122n01}-\Br{A}.
The numbers in square brackets refer to the power of 10.
The values of $p_W$ and $p_{\Asym}$ show the power dependence on $Z$ of $W^{(e)}$ and $\Asym$ ($W^{(e)}\propto Z^{p_W}$, $\Asym\propto Z^{p_{\Asym}}$), respectively.
}
\begin{tabular}{r|ll|lll}
\hline
\multicolumn{1}{c|}{Z}
&\multicolumn{1}{c}{$W^{(e)}$}
&\multicolumn{1}{c|}{$p_W$}
&\multicolumn{1}{c}{$\Asym$}
&\multicolumn{1}{c}{$\Asym^a$}
&\multicolumn{1}{c}{$p_{\Asym}$}\\
\hline
1&8.22906&6.00&4.256617(3)[-6]&4.22[-6]&2.00\\
10&8.20063[6]&5.99&4.27022(1)[-4]&4.24[-4]& 2.01\\
20&5.19515[8]&5.97&1.7242(2)[-3]&1.72[-3]&2.03\\
30&5.82125[9]&5.94&3.9368(3)[-3]&3.97[-3]&2.05\\
40&3.19889[10]&5.90&7.136(2)[-3]&7.32[-3]&2.09\\
50&1.18687[11]&5.84&1.141(1)[-2]&1.20[-2]&2.12\\
60&3.42797[11]&5.78&1.686(2)[-2]&1.84[-2]&2.16\\
64&4.97436[11]&5.75&1.940(3)[-2]&2.16[-1]&2.17\\
70&8.31297[11]&5.70&2.359(4)[-2]&2.71[-2]&2.19\\
80&1.76987[12]&5.60&3.16(1)[-2]&3.92[-2]&2.20\\
90&3.40186[12]&5.47&4.10(2)[-2]&5.62[-2]&2.16\\
92&3.83600[12]&5.44&4.30(2)[-2]&6.06[-2]&2.15\\
100&6.00880[12]&5.30&5.13(3)[-2]&8.19[-2]&2.02\\
110&9.86369[12]&5.07&6.15(5)[-2]&1.23[-1]&1.59\\
118&1.40273[13]&5.02&6.7(1)[-2]&1.80[-1]&0.55\\
120&1.52661[13]&5.12&6.8(2)[-2]&2.00[-1]&0.01\\
\hline
\multicolumn{6}{l}{$\phantom{qqqqq}^{a}$ \cite{au1976}}\\
\end{tabular}
\end{table}

\begin{table}
\caption{\label{tab2}
The differential transition probabilities ($dW^{(e)}/d\omega_1$, in s$^{-1}$keV$^{-1}$)  for two-photon decay of $2s$-electron state and
the asymmetry factor ($\asym(x)$) for $x=1/2$.
The numbers in parentheses indicate the accuracy of the two-parameter approximation
(see
Eqs.~\Br{eqn210121n01y}, \Br{eqn210121n02}).
The numbers in square brackets refer to the power of 10.
The values of $p_W$ and $p_{\asym}$ show the power dependence on $Z$ of $W^{e}$ and $\asym$ ($W^{(e)}\propto Z^{p_W}$, $\asym\propto Z^{p_{\asym}}$), respectively.
}
\begin{tabular}{r|ll|ll}
\hline
\multicolumn{5}{c}{x=1/2}
\\
\hline
\multicolumn{1}{c|}{Z}
&\multicolumn{1}{c}{$dW^{(e)}/d\omega_1$}
&\multicolumn{1}{c|}{$p_W$}
&\multicolumn{1}{c}{$\asym(x)$}
&\multicolumn{1}{c}{$p_{\asym}$}
\\
\hline
1&2.08759[3]&4.00&6.108760(4)[-6]&2.00\\
&&&$<2.5[-4]^a$\\
10&2.08250[7]&4.00&6.118315(4)[-4]&2.00\\
20&3.30725[8]&3.98&2.45890(1)[-3]&2.01\\
30&1.65327[9]&3.95&5.5759(1)[-3]&2.03\\
40&5.13119[9]&3.91&1.00205(3)[-3]&2.05\\
50&1.22281[10]&3.86&1.5872(2)[-2]&2.08\\
54&1.64456[10]&3.83&1.8628(2)[-2]&2.09\\
&&&1.865[-2]$^a$\\
60&2.45826[10]&3.79&2.3229(4)[-2]&2.10\\
64&3.13637[10]&3.75&2.6617(6)[-2]&2.12\\
70&4.38029[10]&3.69&3.220(1)[-2]&2.13\\
80&7.11800[10]&3.56&4.288(3)[-2]&2.16\\
90&1.07276[11]&3.37&5.533(6)[-2]&2.16\\
92&1.15507[11]&3.33&5.801(7)[-2]&2.16\\
&&&5.838[-2]$^a$\\
100&1.51322[11]&3.11&6.93(1)[-2]&2.10\\
110&2.00333[11]&2.71&8.43(2)[-2]&1.91\\
118&2.38972[11]&2.23&9.56(4)[-2]&1.51\\
120&2.47937[11]&2.08&9.70(4)[-2]&1.34\\
\hline
\multicolumn{5}{l}{$\phantom{qqqqq}^{a}$ \cite{surzhykov2005PhysRevA.71.022509}}
\end{tabular}
\end{table}

\begin{table}
\caption{\label{tab3} The same as in Table~\ref{tab2}, but for $x=1/3$.
}
\begin{tabular}{r|ll|ll}
\hline
\multicolumn{5}{c}{x=1/3}
\\
\hline
\multicolumn{1}{c|}{Z}
&\multicolumn{1}{c}{$dW^{(e)}/d\omega_1$}
&\multicolumn{1}{c|}{$p_W$}
&\multicolumn{1}{c}{$\asym(x)$}
&\multicolumn{1}{c}{$p_{\asym}$}
\\
\hline
1&1.98512[3]&4.00&5.12938(2)[-6]&2.00\\
10&1.97912[7]&3.99&5.13841(1)[-4]&2.00\\
20&3.13751[8]&3.97&2.06626(1)[-3]&2.01\\
30&1.56380[9]&3.94&4.6899(1)[-3]&2.03\\
40&4.83349[9]&3.89&8.4387(5)[-3]&2.06\\
50&1.14574[10]&3.83&1.3387(2)[-2]&2.08\\
60&2.28832[10]&3.74&1.962(1)[-2]&2.11\\
64&2.91094[10]&3.70&2.250(1)[-2]&2.13\\
70&4.04597[10]&3.63&2.725(2)[-2]&2.14\\
80&6.51572[10]&3.48&3.633(4)[-2]&2.16\\
90&9.71948[10]&3.27&4.69(1)[-2]&2.15\\
92&1.04422[11]&3.22&4.91(2)[-2]&2.15\\
100&1.35527[11]&2.99&5.87(2)[-2]&2.07\\
110&1.77166[11]&2.56&7.10(3)[-2]&1.81\\
118&2.09131[11]&2.08&7.96(5)[-2]&1.26\\
120&2.16419[11]&1.93&8.14(6)[-2]&1.04\\
\hline
\end{tabular}
\end{table}

\begin{table}
\caption{\label{tab4} The same as in Table~\ref{tab2}, but for $x=1/6$.
}
\begin{tabular}{r|ll|ll}
\hline
\multicolumn{5}{c}{x=1/6}
\\
\hline
\multicolumn{1}{c|}{Z}
&\multicolumn{1}{c}{$dW^{(e)}/d\omega_1$}
&\multicolumn{1}{c|}{$p_W$}
&\multicolumn{1}{c}{$\asym(x)$}
&\multicolumn{1}{c}{$p_{\asym}$}
\\
\hline
1&1.56161[3]&4.00&2.584334(3)[-6]&2.00\\
10&1.55301[7]&3.99&2.591451(4)[-4]&2.01\\
20&2.44352[8]&3.95&1.04514(1)[-3]&2.02\\
30&1.20281[9]&3.90&2.3831(2)[-3]&2.05\\
40&3.65405[9]&3.81&4.313(1)[-3]&2.08\\
50&8.47412[9]&3.71&6.884(4)[-3]&2.11\\
60&1.64860[10]&3.57&1.015(2)[-2]&2.14\\
64&2.07284[10]&3.51&1.165(2)[-2]&2.14\\
70&2.82774[10]&3.40&1.413(3)[-2]&2.14\\
80&4.40119[10]&3.19&1.870(7)[-2]&2.09\\
90&6.32470[10]&2.92&2.40(2)[-2]&1.88\\
92&6.74259[10]&2.85&2.50(2)[-2]&1.88\\
100&8.47504[10]&2.58&2.90(3)[-2]&1.50\\
110&1.06410[11]&2.13&3.23(5)[-2]&0.33\\
118&1.21958[11]&1.70&3.12(7)[-2]&-2.36\\
120&1.25419[11]&1.58&2.98(6)[-2]&-3.70\\
\hline
\end{tabular}
\end{table}

\begin{table*}
\caption{\label{tab5} 
Contributions of the positive and negative energy intermediate states of the electron spectrum to the transition probabilities ($W^{(e)}$, in s$^{-1}$)  for two-photon decay of $2s$ state and
the asymmetry factor ($\Asym$) for one-electron ions.
The columns indicated 'positive' and 'negative' present the results of calculations, where only the positive or negative energy intermediate states are taken into account, respectively.
The notations are the same as in Table \ref{tab1}.
}
\begin{tabular}{r|ll|ll|ll|ll}
\hline
\multicolumn{1}{c|}{}
&\multicolumn{4}{c|}{positive}
&\multicolumn{4}{c}{negative}\\
\hline
\multicolumn{1}{c|}{Z}
&\multicolumn{1}{c}{$W^{(e)}$}
&\multicolumn{1}{c|}{$p_W$}
&\multicolumn{1}{c}{$\Asym$}
&\multicolumn{1}{c|}{$p_{\Asym}$}
&\multicolumn{1}{c}{$W^{(e)}$}
&\multicolumn{1}{c|}{$p_W$}
&\multicolumn{1}{c}{$\Asym$}
&\multicolumn{1}{c}{$p_{\Asym}$}
\\
\hline
1&8.22861&6.00&-6.220120(3)[-7]&2.00&6.25911[-9]&10.00&1.818(3)[-1]&7.49[-4]\\
10&8.15627[6]&5.98&-6.33096(3)[-5]&2.04&6.10945[1]&9.95&1.844(5)[-1]&2.89[-2]\\
50&1.04463[11]&5.59&-2.304(2)[-3]&2.73&4.60068[8]&9.67&2.21(1)[-1]&2.73[-1]\\
92&2.40261[12]&4.25&-1.80(2)[-2]&4.56&1.71783[11]&9.96&2.70(2)[-1]&3.23[-1]\\
120&5.58159[12]&2.18&-8.8(4)[-2]&7.59&2.64389[12]&10.81&2.83(3)[-1]&-7.53[-2]\\
\hline
\end{tabular}
\end{table*}

\begin{table*}
\caption{\label{tab-muon-posneg}
Contributions of the positive and negative energy intermediate states of the muon spectrum to the transition probabilities ($W^{(\mu)}$, in s$^{-1}$)  for two-photon decay of $2s$ state  and
the asymmetry factor ($\Asym$) for one-muon ions.
The columns indicated 'positive' and 'negative' present results of calculations, where only the positive or negative energy intermediate muon states are taken into account, respectively.}
\begin{tabular}{r|ll|ll}
\hline
\multicolumn{1}{c|}{}
&\multicolumn{2}{c|}{positive}
&\multicolumn{2}{c}{negative}\\
\hline
\multicolumn{1}{c|}{Z}
&\multicolumn{1}{c}{$W^{(\mu)}$}
&\multicolumn{1}{c|}{$\Asym$}
&\multicolumn{1}{c}{$W^{(\mu)}$}
&\multicolumn{1}{c}{$\Asym$}
\\
\hline
1
&1.53062[3]&-2.9748(1)[-5]&1.16940[-6]&1.816(5)[-1]\\

10
&2.11820[9]&-4.8(3)[-5]&1.14713[4]&1.89(1)[-1]\\

50
&3.80852[15]&-4.4(3)[-6]&1.02599[10]&3.01(5)[-1]\\

92
&2.08906[17]&-1.9(5)[-6]&2.55222[11]&4.5(1)[-1]\\

120
&6.88561[17]&-1.3(4)[-6]&7.67971[11]&5.5(3)[-1]\\
\hline
\end{tabular}
\end{table*}


\begin{acknowledgments}
The authors are grateful to M. Kaigorodov for providing us with the rms radii of super heavy elements.
The work of V.A.K., K.N.L. and O.Y.A. was supported solely by the Russian Science Foundation under Grant No.~22-12-00043.
The work of K.N.L. was supported by the National Key Research and Development Program of China under Grant No.~2017YFA0402300, the National Natural Science Foundation of China under Grant No.~11774356, and the Chinese Academy of Sciences (CAS) President’s International Fellowship Initiative (PIFI) under Grant No.~2018VMB0016 and Chinese Postdoctoral Science Foundation No.~2020M673538.
\end{acknowledgments}

\end{document}